\documentclass[12pt]{cernprep}
\usepackage{graphicx}
\usepackage{amssymb}
\begin{document}
\newcommand{\dedx}{\mbox{${\rm d}E/{\rm d}x$}}
\newcommand{\EcB}{$E \! \times \! B$}
\newcommand{\omt}{$\omega \tau$}
\newcommand{\omtsq}{$(\omega \tau )^2$}
\newcommand{\rphi}{\mbox{$r \! \cdot \! \phi$}}
\newcommand{\srphi}{\mbox{$\sigma_{r \! \cdot \! \phi}$}}
\newcommand{\dg}{\mbox{`durchgriff'}}
\newcommand{\mg}{\mbox{`margaritka'}}
\newcommand{\pT}{\mbox{$p_{\rm T}$}}
\newcommand{\GeVc}{\mbox{GeV/{\it c}}}
\newcommand{\MeVc}{\mbox{MeV/{\it c}}}
\def\kr{$^{83{\rm m}}$Kr\ }
\begin{titlepage}
\docnum{CERN--PH--EP/2008--022}
\date{4 December 2008}
%
%
\vspace{1cm}
\title{CROSS-SECTIONS OF LARGE-ANGLE HADRON PRODUCTION \\
IN PROTON-- AND PION--NUCLEUS INTERACTIONS I: \\
BERYLLIUM NUCLEI AND BEAM MOMENTA 
OF \mbox{\boldmath $+8.9$}~GeV/\mbox{\boldmath $c$} 
AND \mbox{\boldmath $-8.0$}~GeV/\mbox{\boldmath $c$}}

\begin{abstract}
We report on double-differential inclusive cross-sections of the production of secondary protons, deuterons, and charged pions and kaons, in the interactions with a 5\% $\lambda_{\rm abs}$ thick stationary beryllium target, of a $+8.9$~GeV/{\it c} proton and pion beam, and a $-8.0$~GeV/{\it c} pion beam. Results are given for secondary particles with production angles $20^\circ < \theta < 125^\circ$. 
\end{abstract}

\vfill  \normalsize
\begin{center}
The HARP--CDP group  \\  

\vspace*{2mm} 

A.~Bolshakova$^1$, 
I.~Boyko$^1$, 
G.~Chelkov$^1$, 
D.~Dedovitch$^1$, 
A.~Elagin$^{1a}$, 
M.~Gostkin$^1$,
S.~Grishin$^1$,
A.~Guskov$^1$, 
Z.~Kroumchtein$^1$, 
Yu.~Nefedov$^1$, 
K.~Nikolaev$^1$, 
A.~Zhemchugov$^1$, 
F.~Dydak$^2$, 
J.~Wotschack$^{2*}$, 
A.~De~Min$^{3b}$,
V.~Ammosov$^4$, 
V.~Gapienko$^4$, 
V.~Koreshev$^4$, 
A.~Semak$^4$, 
Yu.~Sviridov$^4$, 
E.~Usenko$^{4c}$, 
V.~Zaets$^4$ 
\\
 
\vspace*{5mm} 

$^1$~{\bf Joint Institute for Nuclear Research, Dubna, Russia} \\
$^2$~{\bf CERN, Geneva, Switzerland} \\ 
$^3$~{\bf Politecnico di Milano and INFN, 
Sezione di Milano-Bicocca, Milan, Italy} \\
$^4$~{\bf Institute of High Energy Physics, Protvino, Russia} \\

\vspace*{5mm}

\submitted{(To be submitted to Eur. Phys. J. C)}
\end{center}

\vspace*{5mm}
\rule{0.9\textwidth}{0.2mm}

\begin{footnotesize}
$^a$~Now at Texas A\&M University, College Station, USA. 

$^b$~On leave of absence at 
Ecole Polytechnique F\'{e}d\'{e}rale, Lausanne, Switzerland. 

$^c$~Now at Institute for Nuclear Research RAS, Moscow, Russia.

$^*$~Corresponding author; e-mail: joerg.wotschack@cern.ch
\end{footnotesize}

\end{titlepage}



\newpage 

\section{Introduction}

\begin{figure}[htp]
\begin{center}
\includegraphics[width=0.8\textwidth]{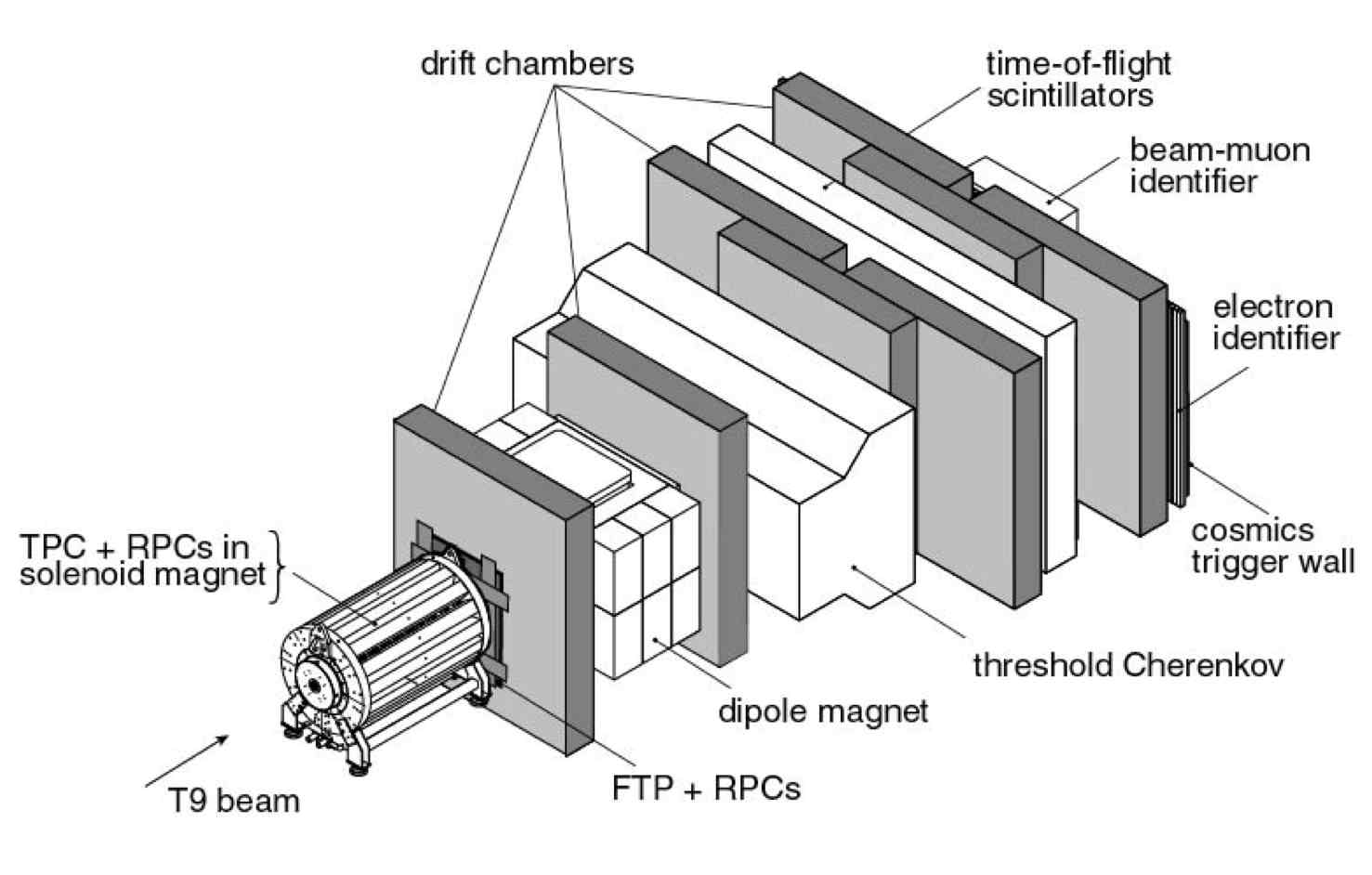} 
\caption{Schematic view of the HARP detector.}
\label{harpdetector}
\end{center}
\end{figure}

The HARP experiment arose from the realization that the 
inclusive differential cross-sections of hadron production 
in the interactions of low-momentum protons with nuclei were 
known only within a factor of two to three, while 
more precise cross-sections are in demand for several reasons. 
Consequently, the HARP detector was designed to carry 
out a programme of systematic and precise measurements of 
hadron production by protons and pions with momenta from 
1.5 to 15~GeV/{\it c}. It is shown schematically in Fig.~\ref{harpdetector}.

The detector extended longitudinally over 14.7~m and combined a 
forward spectrometer with a 
large-angle spectrometer. The latter comprised a 
cylindrical Time Projection 
Chamber (TPC) around the target and an array of 
Resistive Plate Chambers (RPCs) that surrounded the 
TPC. The purpose of the TPC was track 
reconstruction and particle identification by \dedx . The 
purpose of the RPCs was to complement the 
particle identification by time of flight.

The HARP experiment was performed at the CERN Proton Synchrotron 
in 2001 and 2002 with a set of stationary targets ranging 
from hydrogen to lead, including beryllium.

We report on the large-angle production (polar angle $\theta$ in the 
range $20^\circ < \theta < 125^\circ$) 
of secondary protons and charged 
pions, and of deuterons and charged kaons, 
in the interactions with a 5\% $\lambda_{\rm abs}$ Be target
of $+8.9$~GeV/{\it c} protons and pions, and of $-8.0$~GeV/{\it c}
pions. 

The data analysis presented in this paper rests exclusively   
on the calibrations of the TPC and the RPCs that we,
the HARP--CDP group,  
published in Refs.~\cite{TPCpub} and \cite{RPCpub}.  
As discussed in Refs.~\cite{JINSTpub} and \cite{EPJCpub},
and succinctly summarized in this paper's Appendix, 
our calibrations  disagree with calibrations published by the `HARP 
Collaboration'~\cite{HARPTechnicalPaper,OffRPCPaper,
500pseffect,OffTPCcalibration}. 
Conclusions of independent review 
bodies on the discrepancies between our results and those from
the HARP Collaboration can be found in 
Refs.~\cite{CarliFuster,SPSCminutes}.

\section{The T9 proton and pion beams}

The protons and pions were delivered by
the T9 beam line in the East Hall of CERN's Proton Synchrotron.
This beam line supports beam momenta between 1.5 and 15~GeV/{\it c},
with a momentum bite $\Delta p/p \sim 1$\%.

Beam particle identification was provided for by two threshold  
Cherenkov counters filled with nitrogen, and by time of 
flight over a flight path of 24.3~m. In the $+8.9$~GeV/{\it c}
and $-8.0$~GeV/{\it c} beams, the pressure of the
nitrogen gas was set such that protons were below 
threshold for Cherenkov light but pions above. 
The time of flight of each beam particle was measured 
by three scintillation counters
with a precision of 106~ps\footnote{Under stable conditions of 
the beam optics, such that an average 
particle velocity could be used, the time-of-flight precision 
could be improved to 77~ps; in our analysis, no use was made 
of this option.}.  

\begin{figure}[htp]
\begin{center}
\begin{tabular}{cc}
\includegraphics[width=0.5\textwidth]{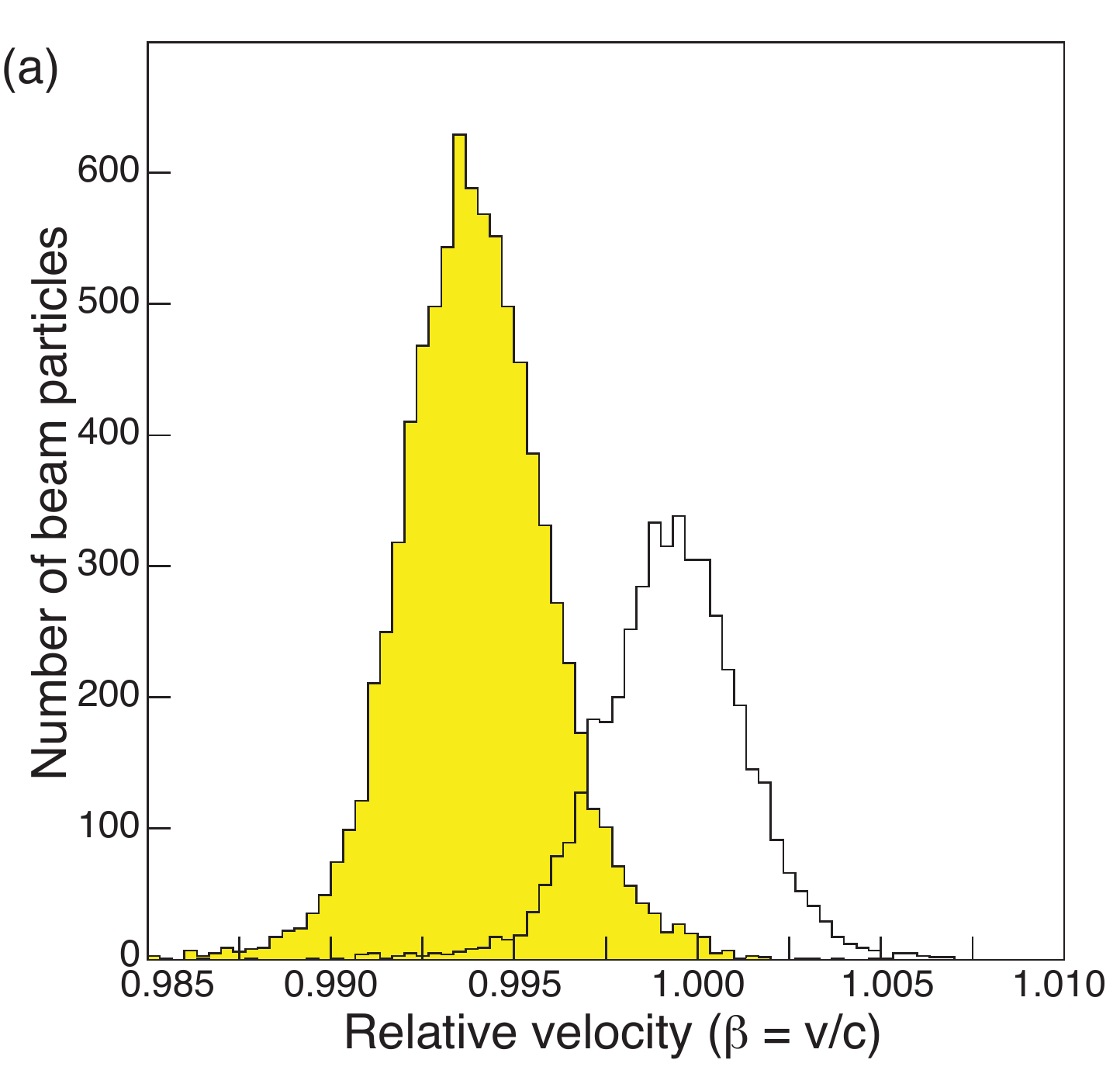} &
\includegraphics[width=0.5\textwidth]{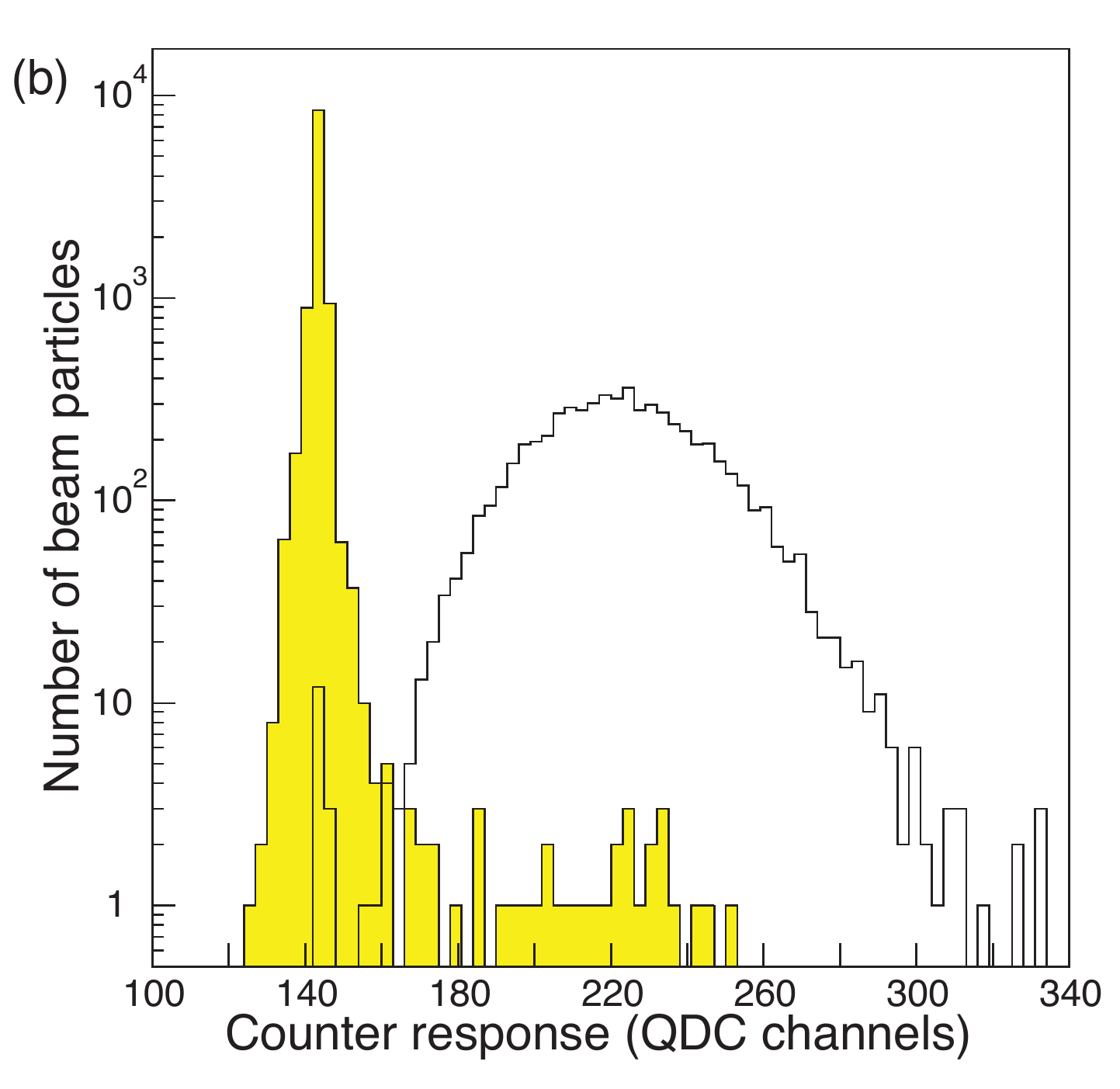}  \\
\end{tabular}
\caption{(a) Relative velocity $\beta$ from the beam time-of-flight 
system of protons (shaded histogram) and pions, with the 
particles identified in the beam Cherenkov counters;
(b) charge response of the light signal in one beam 
Cherenkov counter from 
protons (shaded histogram) and pions, with the particles identified 
in the other beam Cherenkov counter.}
\label{beamparticleid}
\end{center}
\end{figure}

Figure~\ref{beamparticleid}~(a) shows the relative velocity
$\beta$ from the beam time of flight of 
positive particles in the $+8.9$~GeV/{\it c} beam, 
with protons distinguished from `pions'\footnote{The `pions' 
comprise small contaminations by muons and electrons, 
indistinguishable both by time of flight and by beam 
Cherenkov signals.} by the absence of
a beam Cherenkov signal. Vice versa, Fig.~\ref{beamparticleid}~(b)
shows the signal charge in one beam Cherenkov counter, 
with protons and pions distinguished by the signal charge in the
other beam Cherenkov counter. All
measurements are independent of each other and together 
permit a clean separation between protons and pions, respectively, 
with a negligible contamination of less than 0.1\% by the other 
particle.

The pion beam had a contamination by muons from pion decays. 
This contamination was measured to be $(1.7 \pm 0.5)$\%
of the pion component of the $+8.9$~GeV/{\it c}  
beam~\cite{T9beammuons}.
For the $-8.0$~GeV/{\it c} beam, this contamination 
is $(1.9 \pm 0.5)$\%.
The pion beam also had a contamination by electrons from converted
photons from $\pi^0$ decays. This contamination was determined
to be $(1.2 \pm 0.5)$\% of the pion component of the $+8.9$~GeV/{\it c}  
beam~\cite{T9beamelectrons}. We take the same electron fraction 
for the $-8.0$~GeV/{\it c} beam. 
For the determination of interaction cross-sections of pions, 
the muon and electron contaminations must be subtracted from 
the incoming flux of pion-like particles. 

The beam trajectory was determined by a set of three multiwire 
proportional chambers (MWPCs), located upstream of the target,
several metres apart. The transverse error of the 
projected impact point on the target was 0.5~mm from the 
resolution of the MWPCs, plus a
contribution from multiple scattering of the beam particles
in various materials. Excluding the target itself, the 
latter contribution is 0.2~mm for a 8.9~GeV/{\it c} beam 
particle.

The size of the beam spot at the position of the target was several
millimetres in diameter, determined by the setting of the beam
optics and by multiple scattering. 
The nominal beam position\footnote{A 
right-handed Cartesian and/or spherical polar coordinate 
system is employed; the $z$ axis coincides with the beam line, with
$+z$ pointing downstream; the coordinate origin is at the 
centre of the beryllium target, 500~mm
downstream of the TPC's pad plane; 
looking downstream, the $+x$ coordinate points to
the left and the $+y$ coordinate points up; the polar angle
$\theta$ is the angle with respect to the $+z$ axis; when 
looking downstream, the azimuthal angle $\phi$ increases in
the clockwise direction, with the $+x$ axis at $\phi = 0$.} 
was at $x_{\rm beam} = y_{\rm beam} = 0$, however, excursions 
by several millimetres
could occur\footnote{The only relevant issue is that the trajectory
of each individual beam particle is known, whether shifted or not, 
and therefore the amount of matter to be traversed by the 
secondary hadrons.}. 
A loose fiducial cut 
$\sqrt{x^2_{\rm beam} + y^2_{\rm beam}} < 12$~mm
ensured full beam acceptance. The muon and electron 
contaminations of the 
pion beam, stated above, refer to this acceptance cut.

We select  
`good' beam particles by requiring the unambiguous reconstruction
of the particle trajectory with good $\chi^2$. In addition we require that the particle type is unambiguously identified. 
We select `good' accelerator spills by requiring minimal intensity and
a `smooth' variation of beam intensity across the 400~ms long 
spill\footnote{A smooth variation of beam intensity eases 
corrections for dynamic TPC track distortions.}.

\section{The large-angle spectrometer}

In HARP's large-angle region, a cylindrical TPC~\cite{TPCpub} had been chosen
as tracking detector. It was embedded in a solenoidal magnet that generated a magnetic field of 0.7~T parallel to the TPC axis. The magnet was in general operated with its polarity tied to the beam polarity\footnote{The HARP data-taking convention was that $B_{\rm z} > 0$ refers to positive beam polarity.}.
\begin{figure}[htbp]
\begin{center}
\includegraphics[width=\textwidth]{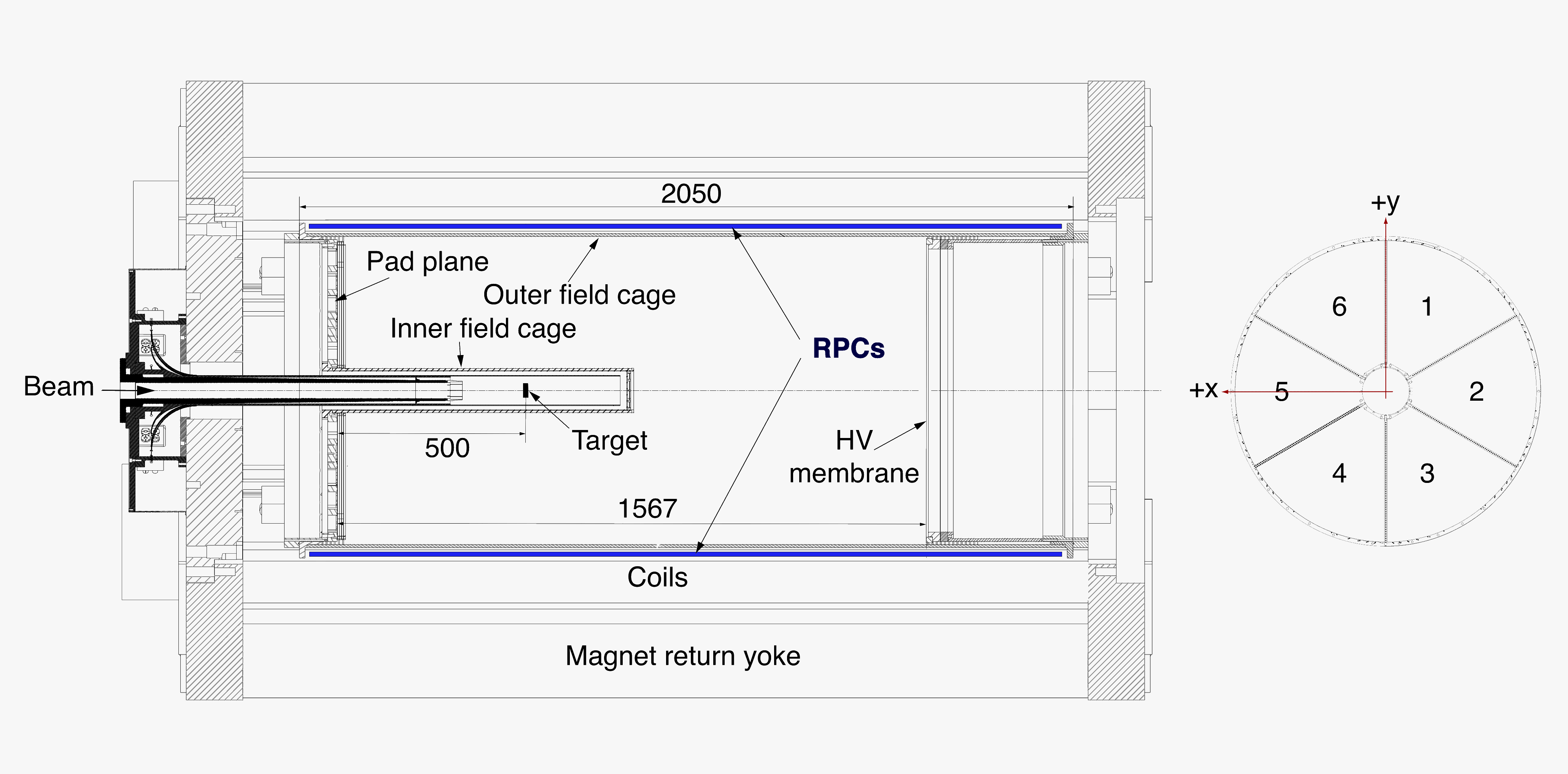}
\caption{Longitudinal cut through the TPC and the solenoidal magnet; the beam enters from the left side; the small figure to the right shows the layout of 
the six TPC readout sectors, looking downstream.}
\label{fig:TPC}
\end{center}
\end{figure} 
 
The TPC filled most of the inner bore of the magnet, leaving a 25~mm wide gap between TPC and magnet coils. This gap was used to house two overlapping layers of 2~m long RPCs~\cite{RPCpub} directly mounted onto the outer field cage of the TPC. 

The layout of the TPC and its position in the solenoidal magnet is shown in Fig.~\ref{fig:TPC}. The TPC has an external diameter of 832~mm and an overall length of $\sim$2~m. It consists of two Stesalit cylinders forming the inner and outer field cages, a wire chamber with pad readout, located at the upstream end, and a high-voltage (HV) membrane at 1567~mm distance from the pad plane. The inner field cage extends over about half of the drift volume; it encloses the target, the centre of which is located 500~mm downstream of the pad plane.

The tracking volume extends radially from 75~mm to 385~mm and over $\sim$1.5~m longitudinally. Electrons from ionization induced by charged particles in the TPC gas drift upstream under the influence of the longitudinal electrical field; they are amplified in the wire chamber and read out through pads arranged in six identical sectors, as shown in Fig.~\ref{fig:TPC}. Each sector comprised 662 readout pads of dimensions $6.5 \times 15$~mm$^2$ arranged in 20 
concentrical rows. 

Our calibration work on the HARP TPC and RPCs is described in Refs.~\cite{TPCpub} and \cite{RPCpub}, and in references cited therein. In particular, we recall that static and dynamic TPC track distortions up to $\sim$10~mm have been corrected to better than 300~$\mu$m. Therefore, TPC track distortions do not affect the precision of our cross-section measurements.  

\section{Track reconstruction and particle identification}

\subsection{Pattern recognition in the TPC}

The clusters measured by the TPC constitute space points 
along the track trajectory. 
Each space point has three uniquely determined 
coordinates: $r$, $\phi$, and $z$.  
Our pattern recognition of tracks with $p_{\rm T} \geq 0.05$~GeV/{\it c} 
originating from the target region 
is based on the TOPAZ histogram 
technique~\cite{TOPAZpatternrecognition}: a 2-dimensional histogram
of the ratio $z/r$ against azimuthal angle $\phi$ is filled 
with all reconstructed clusters. Physical tracks populate
one or two adjacent bins (the bin sizes are suitably
chosen) and thus are easily recognised. 

\subsection{Helix fit of TPC tracks}

For the fit of trajectories in the TPC we adopted the
`Generalized Least-Squares Fit' (GLSF) concept. This is
the formal generalization of the standard 
least-squares fit for an arbitrary number of error-prone dimensions,
and the solution of the equations resulting from the $\chi^2$
minimization with the Lagrange-multiplier method. The 
mathematical intricacies can be found in 
Ref.~\cite{GeneralizedLSF}.    
For the three parameters that describe the circle projection
of a helix, we adopted the
TOPAZ parametrization~\cite{TOPAZ}, for the attractive
feature of avoiding any discontinuity in the numerical
values of fit parameters. Most importantly, it features 
a smooth transition between charge signs of a track.
For more details on the parametrization and
the fit procedure, we refer to Ref.~\cite{GLSFsequel}.

The GLSF must start from reasonable starting values of
the parameters that describe the helix. They are obtained  
by the Chernov--Ososkov least-squares 
algorithm\cite{ChernovOsoskov}. 

Our GLSF algorithm yields the transverse 
momentum $p_{\rm T}$ of a track, its charge sign, its polar angle
$\theta$, and its closest point of approach 
to the $z$ axis.

\subsection{Virtual beam point}

The $p_{\rm T}$ resolution of tracks can be significantly improved by the use of the beam point\footnote{The `beam point' is the best estimate of the
interaction vertex of the incoming beam particle.} as an additional 
point to the trajectory in the TPC. The transverse coordinates of the beam point are known from the extrapolation of the trajectory of the incoming beam particle. Their errors originate from three sources. The first 
is from the extrapolation error of the beam trajectory that is measured by MWPCs; the second stems from multiple scattering of the beam particle; and the third from multiple scattering of the secondary particle in materials between the vertex and the TPC volume.
  
However, the correct error assignment to the beam point is not sufficient. Since a secondary track loses energy by ionization in the target and in materials between the vertex and the TPC volume, a correction must be calculated that replaces the real beam point by a `virtual' beam point which is bias-free with respect to the extrapolation of the trajectory measured in the TPC. It is this virtual beam point, and not the real beam point, that is used in the (final) track fit.
It is determined in an iterative procedure that starts from the fit of the track momentum in the TPC gas, including the real beam point. The fitted trajectory in the TPC gas  is then back-tracked to the beam particle trajectory taking the energy loss and multiple scattering into account. It renders a first estimate of the virtual beam point. Using this estimate, the track in the TPC gas is again fitted, and the procedure is iterated until the position of the virtual beam point is stable.
Since in the calculation of the move from the real to the virtual beam point the energy loss is taken into account, and since the energy loss depends on the type of particle, three different virtual beam points are calculated according to the proton, pion, and electron hypotheses. Accordingly, three different track fits are performed.
 
The fit with the virtual beam point included gives the best possible estimate of the particle momentum in the
TPC gas. In order to determine what is really needed, namely the momentum at the vertex, in a last step the particle is tracked back to the vertex, taking into account the energy loss under the three different particle hypotheses. The track parameters at the vertex are used for the determination of differential cross-sections.

\subsection{Particle identification algorithm}

The particles detected in HARP's large-angle region are
protons, charged pions, and electrons\footnote{The 
term `electron' also refers to positrons.}
(we disregard here small admixtures of kaons and deuterons 
which will be discussed
in Section~\ref{kaonsanddeuterons}).
The charged pion sample comprises muons from pion decay
since the available instrumentation does not distinguish 
them from charged pions.

The \dedx\ and the time-of-flight methods of particle
identification are considered independent. 

To separate measured particles into species, we
assign to each particle a probability of being a proton,
a pion (muon), or an electron, respectively. The probabilities
add up to unity, so that the number of particles is conserved.

Each track is characterized by four measured quantities: $p_{\rm T}$ (transverse momentum), $\theta$ (polar angle), $\beta$ (relative velocity) and \dedx\ (specific ionization).
For particle identification purposes, these variables refer
to reconstructed (`smeared') variables in both the data and 
the Monte Carlo simulation.
 
In every bin of $(p_{\rm T},\theta)$, the probability 
${\cal P} (i|\beta, {\rm d}E/{\rm d}x, p_{\rm T},\theta)$
of a particle to belong to species $i$ ($i$~=~1 [proton], 2 [pion], 
3 [electron]) in a mixture of protons, pions, and electrons  
is according to Bayes' theorem as follows:
\begin{equation}
{\cal P}(i|\beta,{\rm d}E/{\rm d}x,p_{\rm T},\theta) =
   \frac{P(\beta,{\rm d}E/{\rm d}x | i,p_{\rm T},\theta)  
   \cdot P(i,p_{\rm T},\theta)}
   {\sum_{i=1}^3 \left[ P(\beta,{\rm d}E/{\rm d}x | i,p_{\rm T},\theta)
   \cdot P(i,p_{\rm T},\theta) \right] }   \; ,
\label{bayes1}
\end{equation} 
where the sum 
\begin{displaymath}
\sum_{i=1}^3 {\cal P} (i|\beta,{\rm d}E/{\rm d}x,p_{\rm T},\theta)
\end{displaymath}
is normalized to unity. The probabilities $P(i,p_{\rm T},\theta)$
are given by
\begin{displaymath}
P(i,p_{\rm T},\theta) =
   \frac{N_i (p_{\rm T},\theta)}
   {\sum_{i=1}^3 N_i (p_{\rm T},\theta)}  \; ,
\end{displaymath}
where $N_i (p_{\rm T},\theta)$ is the number of particles of species
$i$ in the respective data sample.
Then Eq.~(\ref{bayes1}) becomes
\begin{equation}
{\cal P} (i|\beta,{\rm d}E/{\rm d}x,p_{\rm T},\theta) =
   \frac{P(\beta,{\rm d}E/{\rm d}x | i,p_{\rm T},\theta)
   \cdot N_i (p_{\rm T},\theta)}
   {\sum_i^3 \left[ P(\beta,{\rm d}E/{\rm d}x | i,p_{\rm T},\theta)
   \cdot N_i (p_{\rm T},\theta) \right] }  \; .
\label{bayes2}
\end{equation} 
We note that in Eqs.~(\ref{bayes1}) and (\ref{bayes2}) the
term $P(\beta,{\rm d}E/{\rm d}x|i,p_{\rm T},\theta)$ denotes
a probability density function which is normalized to unity.
This probability density function must represent the data
in the bin $(p_{\rm T},\theta)$.

Before determining the probability represented by Eq.~(\ref{bayes2}),
the probability density functions
$P(\beta,{\rm d}E/{\rm d}x|i,p_{\rm T},\theta)$ and
the particle abundances $N_i (p_{\rm T},\theta)$ must be known.
This seemingly circular situation is resolved by an iterative 
comparison of data with the Monte Carlo simulation, to achieve  
agreement of the distributions in both 
variables $\beta$ and \dedx . With a view to starting
from abundances as realistic as possible, the comparison
is initially limited to regions in phase space where the particle species
are unambiguously separated from each other in either \dedx\ or
$\beta$. In other words, the few parameters  
that govern the probability density functions and the 
particle abundances are determined from the data in every bin of  
$(p_{\rm T},\theta)$.

In case one of the two identification
variables is absent\footnote{For example, because of too few 
clusters to calculate \dedx , or a missing RPC pad.}, only the 
other is used. In the rare cases where
both identification variables are absent, the identification
probabilities reproduce the estimated particle abundances.

\subsection{Particle abundances}

Particle abundances cannot {\it a priori\/} be expected to be
correct in the Monte Carlo simulation. Therefore in general the
particles must be weighted such that data and Monte Carlo
distributions agree.  

We had expected
that the Monte Carlo simulation tool kit
Geant4~\cite{Geant4} would provide us with 
reasonably realistic spectra of secondary hadrons. 
We found this expectation more or less  
met by Geant4's so-called QGSP\_BIC physics list, but only
for the secondaries from incoming beam protons.
For the
secondaries from incoming beam pions, we found the standard 
physics lists of Geant4 unsuitable~\cite{GEANTpub}. 

To overcome this problem,
we built our own HARP\_CDP physics list
for the production of secondaries from incoming beam pions. 
It starts from Geant4's standard QBBC physics list, 
but the Quark--Gluon String Model is replaced by the 
FRITIOF string fragmentation model for
kinetic energy $E>6$~GeV; for $E<6$~GeV, the Bertini 
Cascade is used for pions, and the Binary Cascade for protons; 
elastic and quasi-elastic scattering is disabled.
Examples of the good performance of the HARP\_CDP physics list
are given in Ref.~\cite{GEANTpub}.

\begin{figure}[h]
\begin{center}
\begin{tabular}{c}
\includegraphics[width=0.7\textwidth]{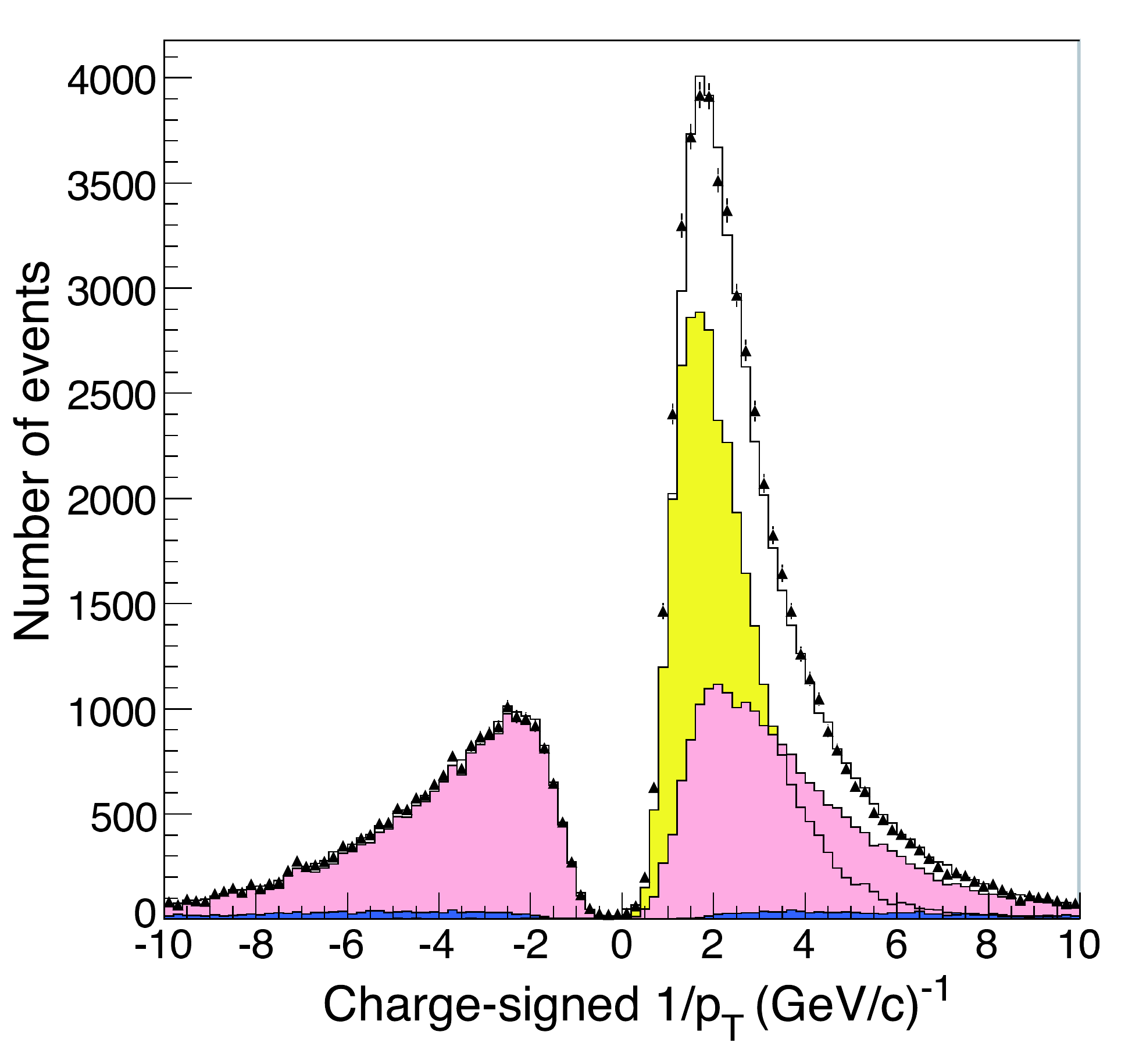} 
\end{tabular}
\end{center}
\caption{$1/p_{\rm T}$ spectra of the secondary particles
from $+8.9$~GeV/{\it c} beam protons on 
a 5\% $\lambda_{\rm abs}$ Be target, for polar angles
$50^\circ < \theta < 60^\circ$; black triangles denote data, the solid
lines Monte Carlo simulation; the shaded histograms show the 
subdivision of the 
data into particle species by applying the particle
identification weights: light shading denotes protons, medium shading
pions, and dark shading electrons.}
\label{abundances}
\end{figure}

Figure~\ref{abundances} demonstrates the level of overall agreement 
between data and Monte Carlo simulation in the variable $1/p_{\rm T}$, 
after convergence of the 
iterative procedure to determine the smooth weighting functions
to the latter.  
The figure also shows, for incoming protons and for a typical
polar-angle range, the subdivision of the 
data into particle species by applying the particle
identification weights. 

Once the abundances are determined, for any pair of \dedx\ and
$\beta$, and using the experimental
resolution functions, the probability can be derived that the particle is 
a proton, a pion, or an electron. This probability is
consistently used for weighting when entering tracks into
plots or tables.

\section{Physics performance}
\subsection{Physics performance of the TPC}

From the requirement that a
$\pi^+$ and a $\pi^-$ with the same RPC time of flight  
have the same momentum, and from the error of the 
magnetic field strength which is less than 1\%, 
the absolute momentum scale is
determined to be correct to 
better than 2\%, both for positively and negatively
charged particles.

\begin{figure}[htp]
\begin{center}
\includegraphics[width=0.95\textwidth]{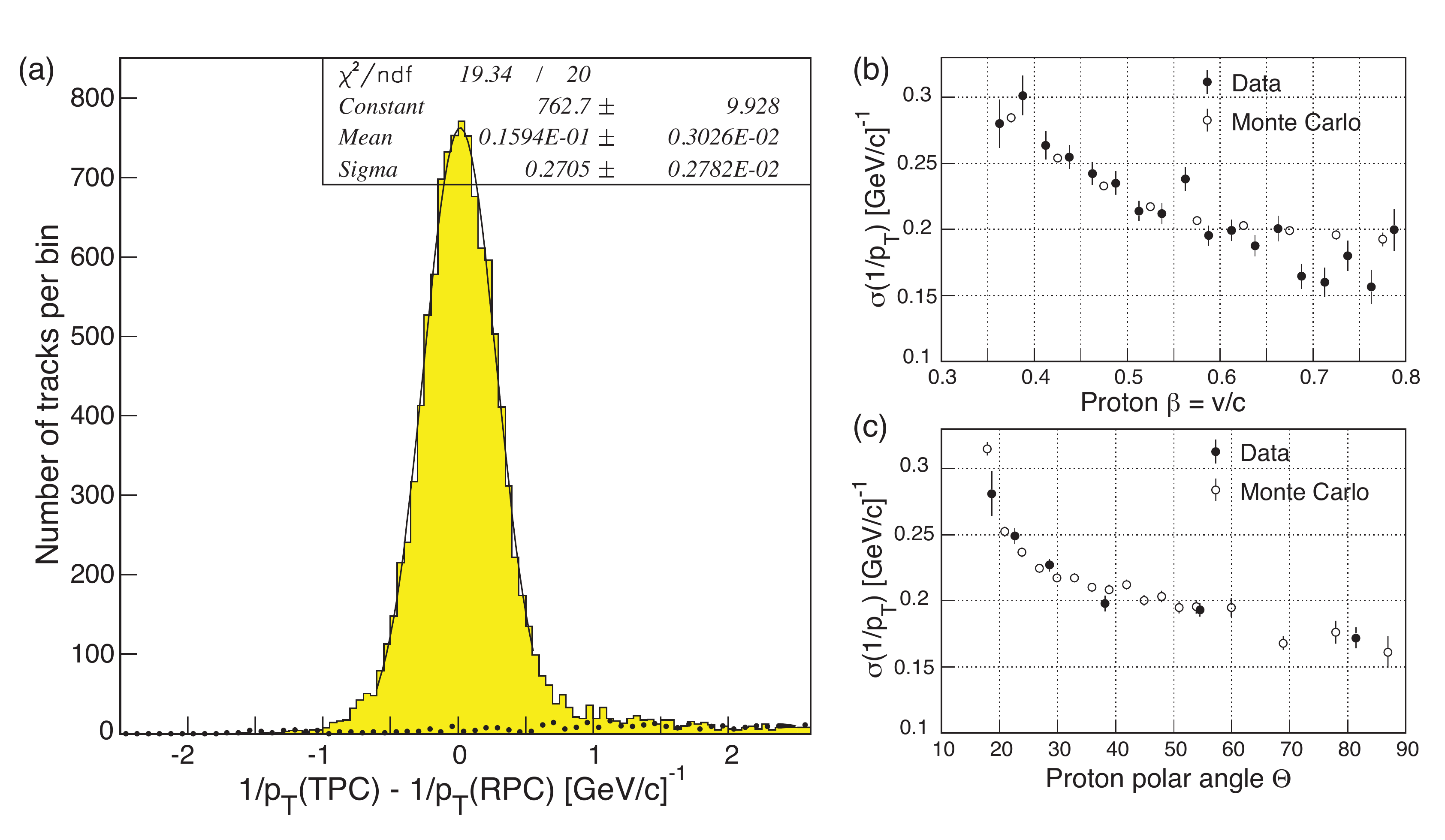}  
\caption{(a) Difference of the inverse transverse momenta of positive
(shaded histogram) and negative (dots) particles 
from the measurement in the TPC and from the determination from 
RPC time of flight, for $0.6 < \beta < 0.75$ and 
for $45^\circ < \theta < 65^\circ$; the positive
particles are protons, with a very small background from pions
and kaons; (b) $\sigma (1/p_{\rm T})$ of protons with 
$45^\circ < \theta < 65^\circ$ as a function
of their relative velocity $\beta$; (c) $\sigma (1/p_{\rm T})$ of protons 
with $0.6 < \beta < 0.75$ as a function of their polar angle 
$\theta$.}
\label{pTresolutionphysics}
\end{center}
\end{figure}

Figure~\ref{pTresolutionphysics}~(a) shows the $1/p_{\rm T}$ 
difference for positive particles with
$0.6 < \beta < 0.75$ and $45^\circ < \theta < 65^\circ$, 
between the measurement in the TPC and the
determination from RPC time of flight with the proton-mass
hypothesis. The selection cuts ensure a practically pure 
sample of protons (the background from pions and kaons  
is negligible as suggested by the very small contribution
of negative particles selected with the same cuts that are shown 
as dots in Fig.~\ref{pTresolutionphysics}~(a)).
A net TPC resolution of
$\sigma (1/p_{\rm T}) = 0.20$~(GeV/{\it c})$^{-1}$
is obtained by subtracting the  
contribution of $\sim$0.18~(GeV/{\it c})$^{-1}$ from the 
time-of-flight resolution and fluctuations from
energy loss and multiple scattering in materials
between the vertex and the TPC volume 
quadratically from the convoluted 
resolution of 0.27~(GeV/{\it c})$^{-1}$. 

Figure~\ref{pTresolutionphysics}~(b) shows the net TPC resolution
$\sigma (1/p_{\rm T})$ as a function 
of $\beta$. Figure~\ref{pTresolutionphysics}~(c) shows
the same as a function of $\theta$. The
agreement with the expectation from a Monte Carlo simulation
is satisfactory. The resolution $\sigma (1/p_{\rm T})$
is typically 20\% and worsens towards small~$\beta$ and
small~$\theta$. This is because in
both cases the position error of the virtual beam point 
increases owing to increased multiple scattering 
in materials before the protons enter the TPC.

\begin{figure}[htp]
\begin{center}
\includegraphics[width=0.6\textwidth]{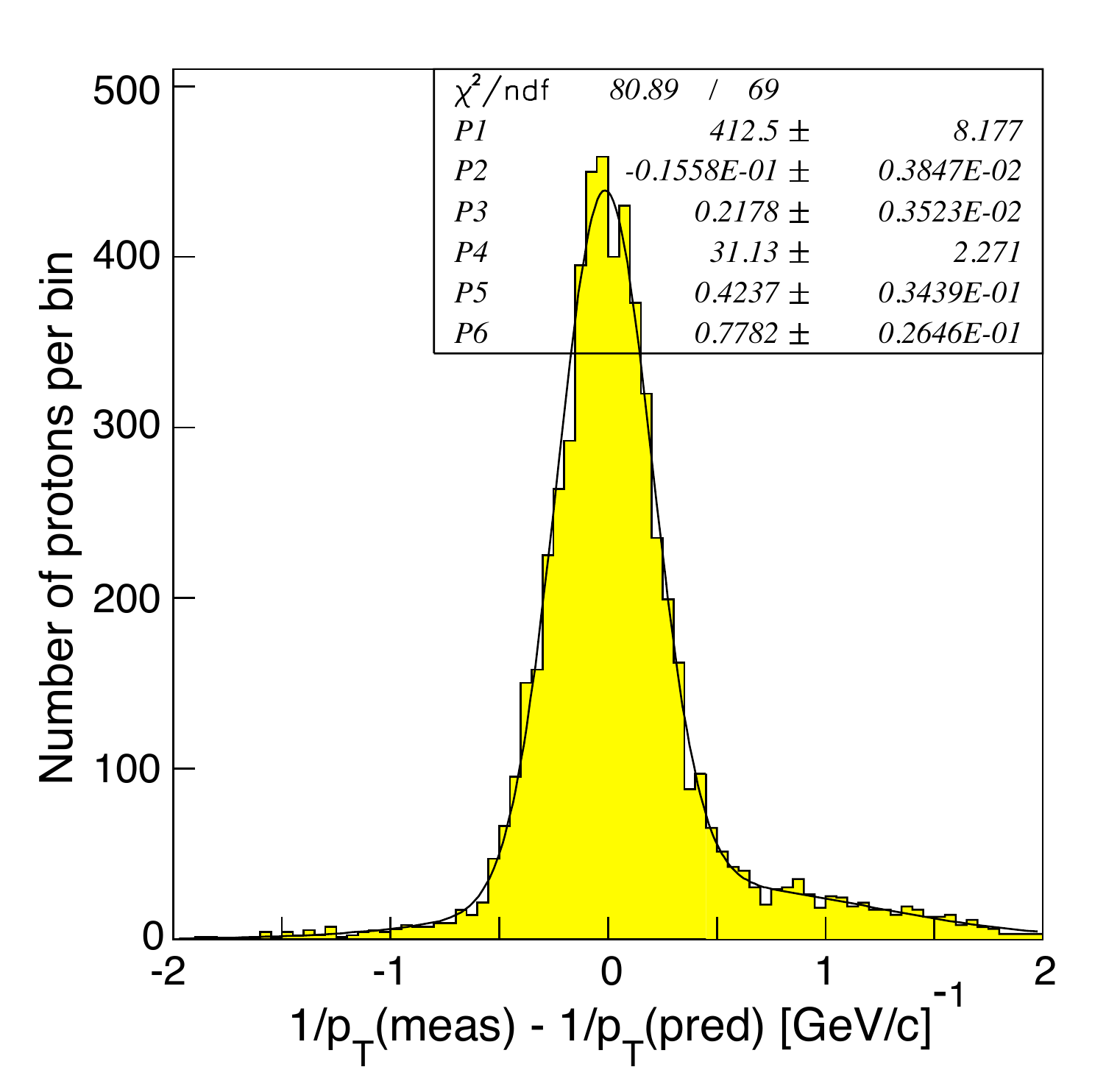} 
\caption{The difference $1/p^{\rm meas}_{\rm T} - 1/p^{\rm pred}_{\rm T}$ 
from large-angle recoil protons
in elastic scattering events from a  
$+3$~GeV/{\it c} beam impinging on a liquid 
hydrogen target; the tail at the right side reflects the
Landau tail in the proton energy loss in materials between the
interaction point and the TPC volume.}
\label{hydrogen}
\end{center}
\end{figure}

Data from the elastic scattering of incoming pions or protons 
on protons at rest have the
added feature of a kinematical constraint. The possibility
to calculate from the four-momentum of the incoming beam particle
and the polar angle $\theta$ the momentum of
the large-angle recoil proton, permits a valuable cross-check
of the TPC's $p_{\rm T}$ resolution.  
Figure~\ref{hydrogen} shows the result from 
the elastic scattering of incoming $+3$~GeV/{\it c} protons 
and $\pi^+$'s in a liquid hydrogen target.  
Here, the $p_{\rm T}$ of the recoil proton has been determined in 
the following two ways:
$1/p^{\rm meas}_{\rm T}$ is determined from the reconstructed 
track curvature in the TPC;  
$1/p^{\rm pred}_{\rm T}$ is predicted from the
elastic scattering kinematics from the polar angle of the recoil
proton which is little affected by TPC track distortions.
Figure~\ref{hydrogen} 
demonstrates a resolution 
of $\sigma(1/p_{\rm T}) \sim 0.19$~(GeV/{\it c})$^{-1}$ after 
unfolding a contribution
of $\sigma (1/p_{\rm T}) \sim 0.12$~(GeV/{\it c})$^{-1}$ from 
fluctuations from energy loss and multiple scattering in materials
between the vertex and the TPC volume.
The measured difference in $p_{\rm T}$ is 0.8\%, in line with the 2\% uncertainty of the momentum scale.  

The polar angle $\theta$ is measured in the TPC with a 
resolution of $\sim$9~mrad, for a representative 
angle of $\theta = 60^\circ$. To this a multiple scattering
error has to be added which is $\sim$7~mrad for a proton with 
$p_{\rm T} = 500$~MeV/{\it c} and $\theta = 60^\circ$, and 
$\sim$4~mrad for a pion with the same characteristics.
The polar-angle scale is correct to better than 2~mrad.     

Besides the $p_{\rm T}$ and the polar angle $\theta$ of tracks,
the TPC also measures \dedx\ with a view to particle identification. 
The \dedx\ resolution is 16\% for a track length of 300~mm.

\subsection{Physics performance of the RPCs}

The intrinsic efficiency of the RPCs that surround 
the TPC is better than 98\%. While
the system efficiency for pions with $p_{\rm T} > 100$~MeV/{\it c} 
at the vertex is close to the intrinsic efficiency, it is 
slightly worse for protons because of their higher energy loss in
structural materials. Protons with $p < 350$~MeV/{\it c}
at the vertex get absorbed before they reach the
RPCs and thus escape time-of-flight measurement.
 
The intrinsic time resolution of the RPCs is 127~ps and
the system time-of-flight resolution (that includes the
jitter of the arrival time of the beam particle at the target)
is 175~ps.  

\begin{figure}[h]
\begin{center}
\begin{tabular}{cc} 
\includegraphics[height=0.4\textwidth]{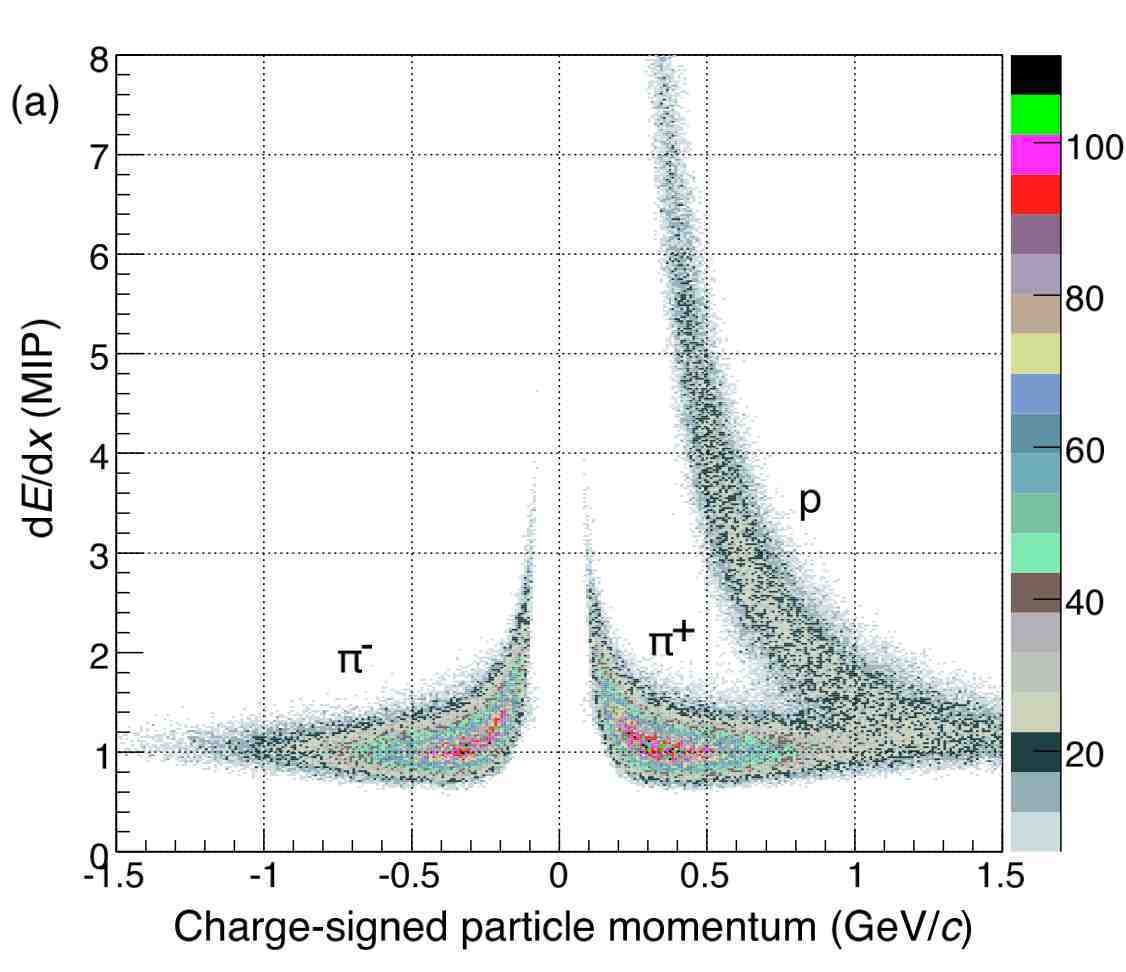} &
\includegraphics[height=0.4\textwidth]{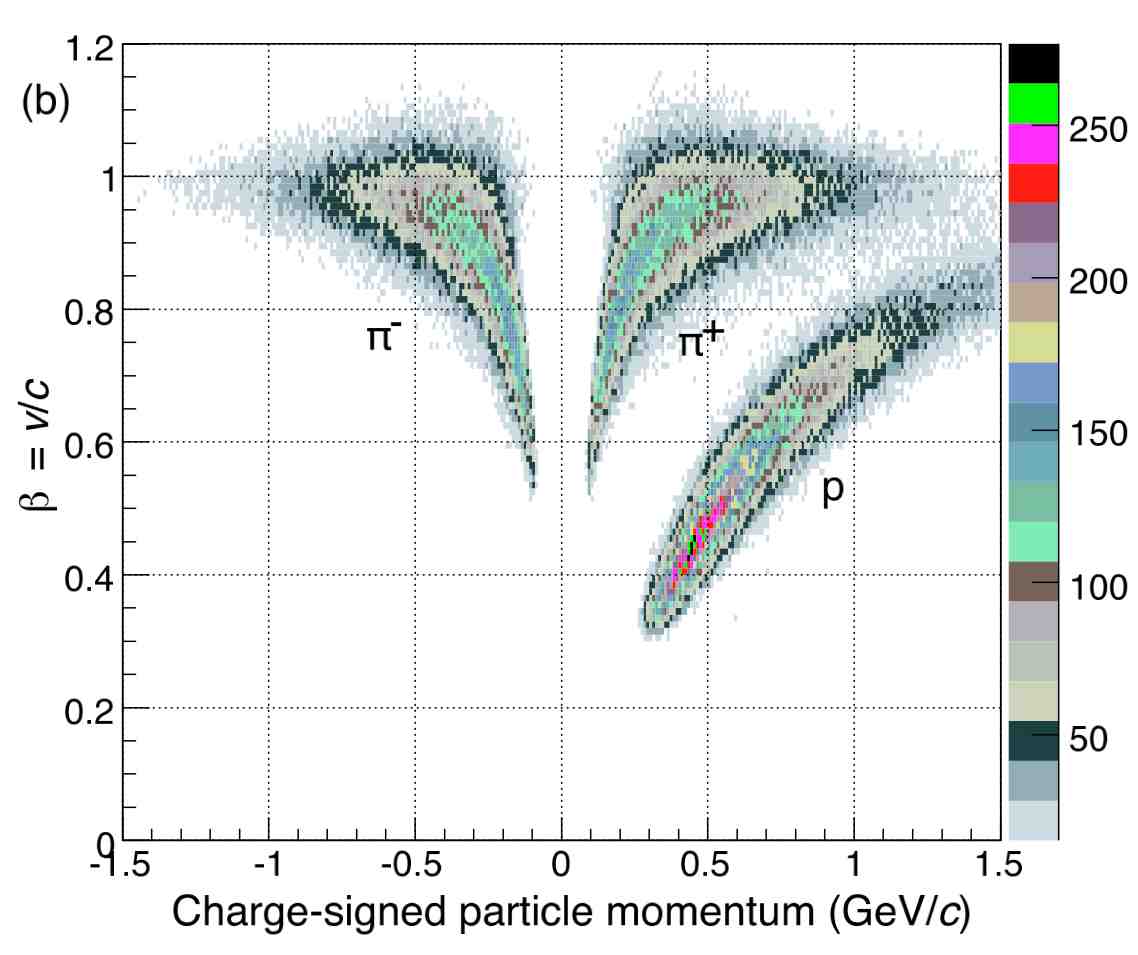} \\
\end{tabular}
\end{center}
\caption{(a) Specific ionization d$E$/d$x$ and (b) velocity $\beta$, 
versus the charge-signed momentum of positive and negative tracks 
in $+8.9$~GeV/{\it c} data; the boxes at the right side indicate 
the event statistics.}
\label{dedxandbeta}
\end{figure}

Figure~\ref{dedxandbeta}~(a) shows the specific ionization
\dedx , measured by the TPC, and Fig.~\ref{dedxandbeta}~(b) the 
relative velocity $\beta$ 
from the RPC time of flight, of positive and negative secondaries, 
as a function of the momentum measured in the TPC. The figures 
demonstrate that in general protons and pions are well separated.
They also underline the importance of the complementary 
separation by RPC time of flight at large particle momentum. 

\section{Normalized secondary particle flux}

The measurement of the inclusive double-differential 
cross-section ${\rm d}^2 \sigma / {\rm d} p {\rm d} \Omega$ 
requires the flux of incoming beam particles, the number
of target nuclei, and the number of secondary particles
in bins of momentum $p$ and polar angle $\theta$. We shall 
discuss these elements in turn. 

\subsection{Beam intensity}

The event trigger had two levels. A first, loose, level required 
only time-coincident signals from beam scintillation counters.
Irrespective of an interaction in the target, each 64th
coincidence signal requested data readout as `beam trigger'. 
A second, tighter level required in addition a signal in a 
cylindrical scintillator fibre detector that surrounded 
the target region, or a signal in a plane of scintillators in
the forward direction (termed `FTP' in Fig.~\ref{fig:TPC}). 
Each such `event trigger'   
also requested data readout.

To achieve the wanted event statistics, the 
experiment was typically run with a dead time in excess of 50\%,
given the 400~ms long 
accelerator spill and a readout time of order 1~ms per event.
Since the dead time affects the beam trigger and the event
trigger in the same way, it cancels in the
cross-section calculation. For a given data set,
the flux of incoming beam particles is defined by the 
number of beam triggers, multiplied by the scale-down
factor of 64. It is imperative, though, that 
the same cuts on the quality of the trajectory of the
beam particle and on its identification be applied
for accepted beam triggers and for accepted 
event triggers. 

The efficiencies of both the beam trigger and the event trigger
are very close to 100\%, thanks to majority requirements.
The beam trigger efficiency cancels. For the event trigger,
we determined an efficiency of $(99.0 \pm 0.2)$\%.
   
\subsection{Target}

The target was a cylinder made of 
high-purity (99.95\%) beryllium, with a density of 1.85~g/cm$^3$,
a radius of 15~mm, and a thickness of $20.5 \pm 0.1$~mm 
(5\% $\lambda_{\rm abs}$).

The finite thickness of the target leads to a
small attenuation of the number of incident beam particles. The
attenuation factor is $f_{\rm att} = 0.975$.

\subsection{Track counting in bins of $p_{\rm T}$ and $\theta$}

This paper is concerned with determining inclusive 
cross-sections of secondaries from the interactions of protons
and pions with beryllium nuclei. This means that for a given
data set, the secondaries are weighted with their probability 
of being a proton, a pion, or an electron, 
counted in bins of $p_{\rm T}$ and $\theta$, and related to 
the number of incoming beam particles and the number of
target nuclei. The counting of secondaries 
is done in an integral way without regard to track--event 
relations.

Electrons stem primarily from the 
conversion of photons from $\pi^0$ decays. They tend to 
concentrate at small momenta. Below 150~MeV/{\it c}, they
are identified by both \dedx\ and time of flight from the
RPCs. From 150 to 250~MeV/{\it c}, the \dedx\ of pions and 
electrons coincides and they are only identified 
by time of flight. The Geant4 electron abundance is compared 
with data in the region of good separation, as a function 
of momentum, and weighted to agree with the data. 
In the region of bad separation the electrons are subtracted
using not the electron abundance predicted by Geant4, 
but the weighted 
prediction extrapolated from the region of good separation. 
Therefore, the Geant4 prediction is used only through its
extrapolated prediction of the energy dependence of 
electrons with momentum larger than 250~MeV/{\it c}. 

Since the particle identification algorithm 
assigns to every particle a probability of being a proton, a
pion, or an electron, the elimination of electrons from
the samples of secondary protons and pions is straightforward.  

It is justified to think of secondary tracks as originating 
exclusively from proton and pion interactions: interactions 
of beam electrons might occasionally lead to low-momentum 
electron or positron tracks in the TPC, however, such tracks 
are recognized by the particle identification algorithm
and disregarded in hadron production cross-sections.
Interactions of beam muons can be neglected.

Kaon and deuteron secondaries are initially part of 
pions and protons, respectively. Their identification is
dealt with in Section~\ref{kaonsanddeuterons}.  

\subsection{Track selection cuts}

We have a selection of `good' TPC sectors: we discard tracks from the `horizontal' sectors 2 and 5 out of the six sectors (see Fig.~\ref{fig:TPC}) for reasons of much worse than average performance, and the lack of reliable track 
distortion corrections~\cite{TPCpub}. 

Tracks are accepted if there are at least 10 TPC clusters along the trajectory. 

A cut in the azimuthal angle $\phi$ is applied to avoid
the dead regions of the six `spokes' that subdivide the TPC pad
plane into six sectors: 10$^\circ$ on one side and 2$^\circ$ on
the other side of each spoke for tracks of one charge, and 
{\it vice versa\/} for the other charge. The asymmetric cut is 
motivated by the opposite bending of positive and negative 
tracks in the magnetic field.

The polar-angle range of tracks is limited to the range 
$20^\circ < \theta < 125^\circ$. Tracks are also required to 
point back to the target, within the resolution limits.

\subsection{Correction for inefficiencies of track reconstruction and track selection}

\begin{figure}[ht]
\begin{center}
\vspace*{3mm}
\begin{tabular}{ccc}
\includegraphics[width=0.3\textwidth]{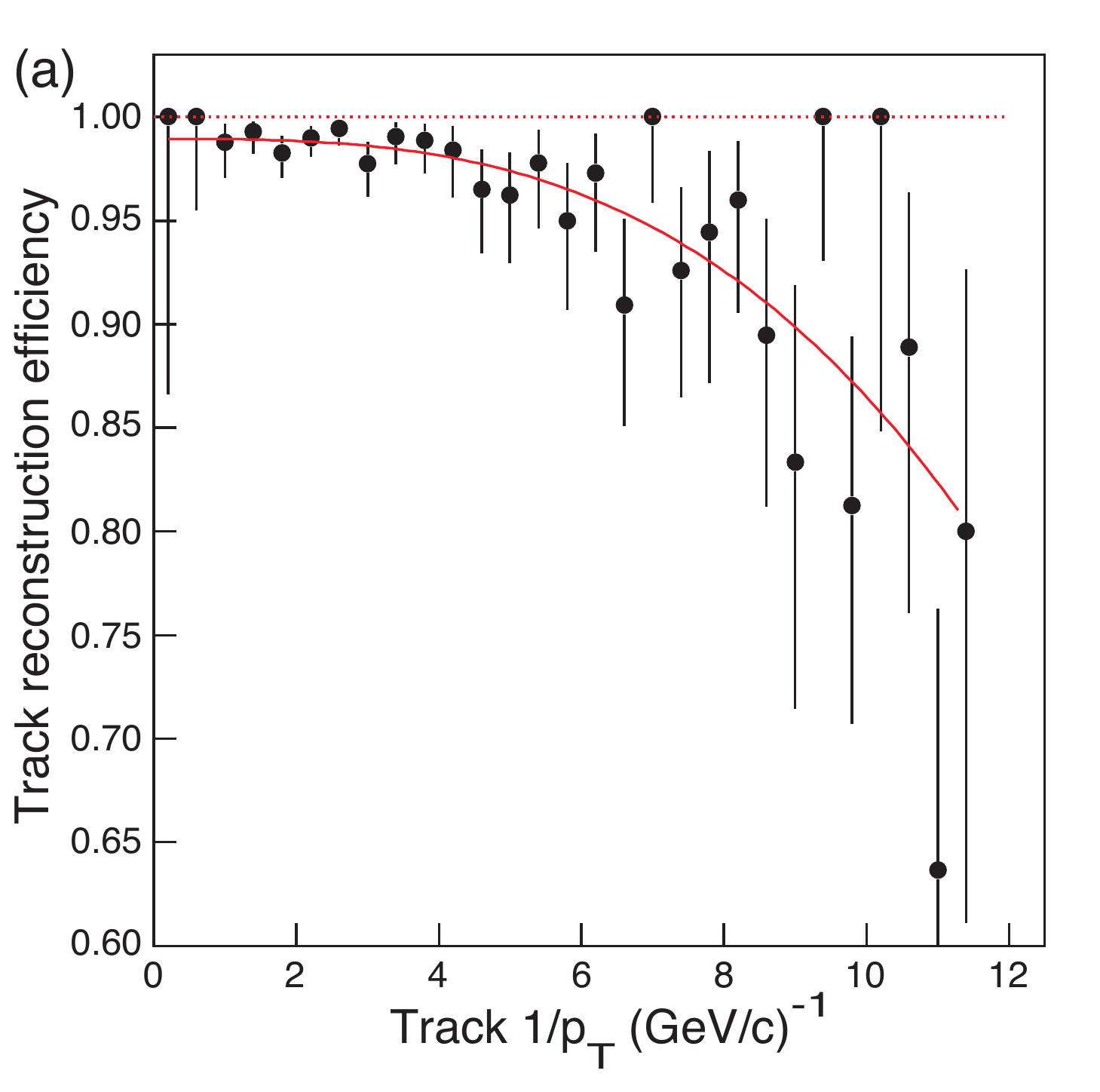} &
\includegraphics[width=0.3\textwidth]{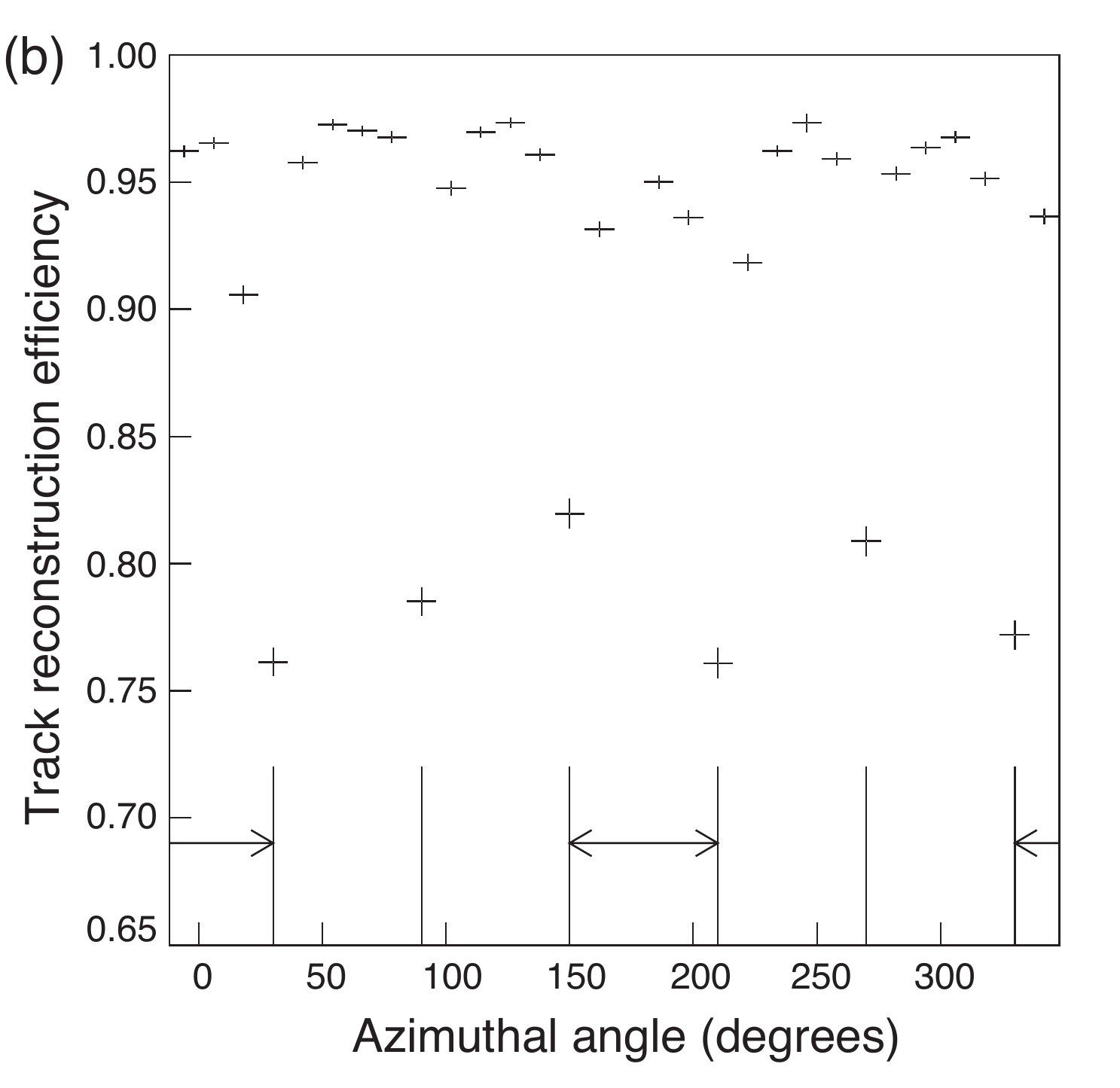} &
\includegraphics[width=0.305\textwidth]{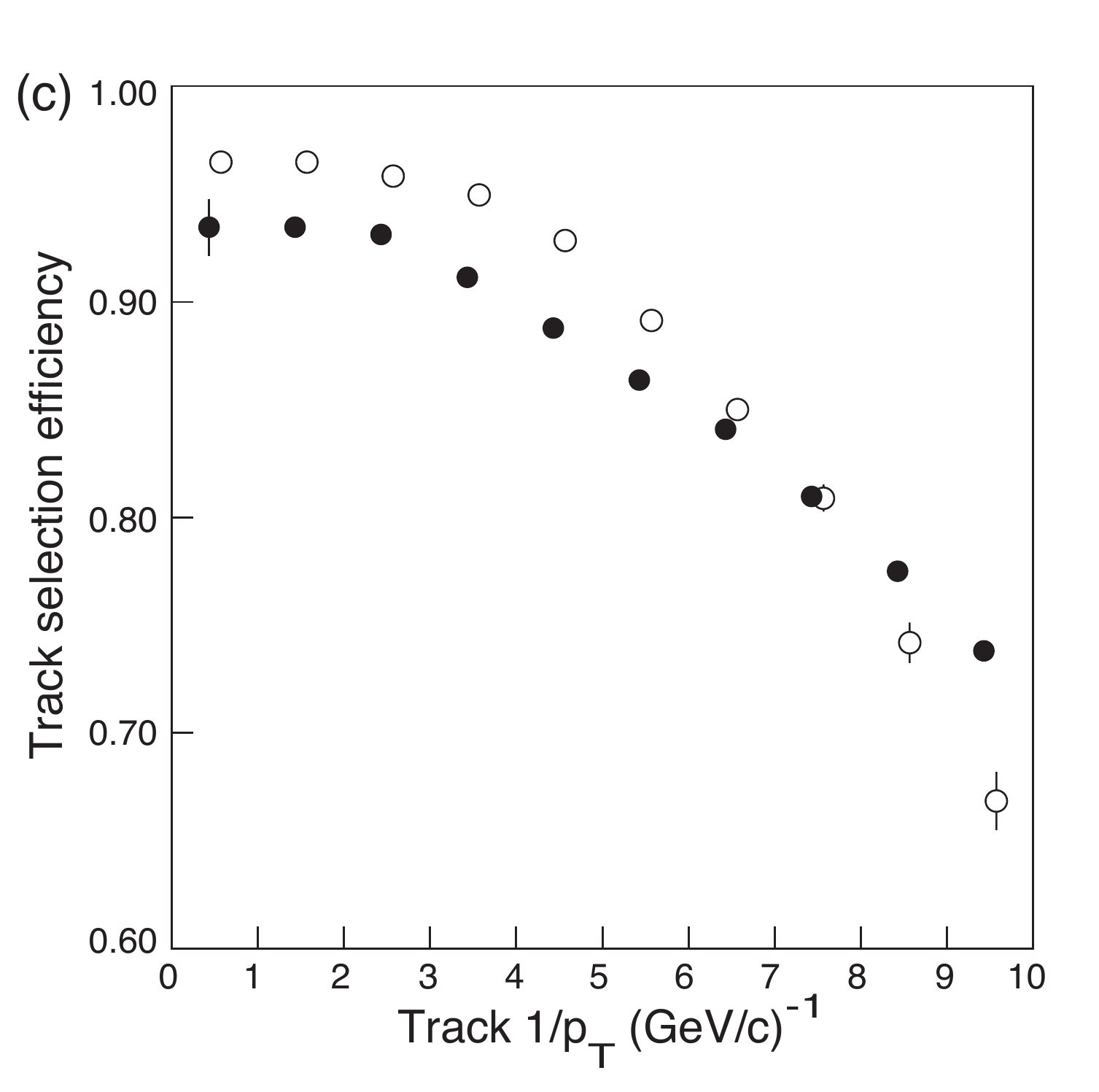} \\
\end{tabular}
\caption{(a) Track reconstruction efficiency from the eyeball scan; (b) track reconstruction efficiency from RPC hits; the vertical lines denote TPC sector boundaries where the TPC `spokes' render the efficiency lower; the azimuthal ranges of the two TPC sectors not used for the analysis are indicated by arrows; (c) efficiency of 10 or more TPC clusters for protons (open circles) and for pions (black circles); the errors shown are statistical only and mostly smaller than 
the symbol size.}
\label{reconstructionefficiency}
\end{center}
\end{figure}

The track reconstruction efficiency was determined by eyeball
scanning of several thousand events by several physicists,
with consistent results among them. The large number of
scanned events permits us to determine the reconstruction
efficiency as a function of geometric or kinematic variables.
For example, Fig.~\ref{reconstructionefficiency}~(a) shows the reconstruction efficiency as a function of $1/p_{\rm T}$ for all cases where the human eye finds at least five (out of a maximum of 20) clusters along a trajectory.
The average reconstruction efficiency is between 95\% and 97\%\footnote{This average holds for tracks with $|p_{\rm T}| > 0.1$~GeV/{\it c} and $20^\circ < \theta < 125^\circ$, with safe distance from the insensitive azimuthal regions caused by the TPC `spokes'.}, where the 2\% range reflects the variation between different data sets.

We cross-checked the track reconstruction efficiency by requiring an RPC hit and at least two TPC clusters in the cone that is subtended by the respective RPC pad, as seen from the vertex. Figure~\ref{reconstructionefficiency}~(b) shows the resulting reconstruction efficiency as a function of the track's azimuthal angle. Outside the TPC spokes and within the four `good' TPC sectors, the reconstruction efficiency determined that way agrees with the result from the eyeball scan.

The requirement of a minimum of 10 TPC clusters per track entails a loss that must be accounted for.  
Since the TPC cluster charge is in general larger for
protons\footnote{The lower proton velocity leads to 
higher specific ionization.} than for pions,
the loss from this cut is different for protons and 
pions. Figure~\ref{reconstructionefficiency}~(c) shows
the efficiency of requiring 10 or more TPC clusters
as a function of $1/p_{\rm T}$, separately for protons
and pions (the average number of clusters was $\sim$14).

The overall track efficiency was taken as the product of the
track reconstruction efficiency and the probability of
having at least 10 clusters along the trajectory.

\subsection{Further corrections}

In this section, we discuss a few more corrections that are
applied to the data. In general, they are determined from a 
Monte Carlo simulation that reproduces the migration
of track parameters from generated (`true') to reconstructed 
(`smeared') ones. This concerns effects arising from finite resolution, 
charge misidentification, pion decays into muons, and 
re-interactions of secondaries in materials between the vertex 
and the TPC volume\footnote{These re-interactions, especially 
in the target material, are different from re-interactions of 
secondaries in the nuclear matter of the same nucleus in 
which the incoming beam particle interacted; the latter 
is an integral part of the inclusive cross-section  
reported in this paper.}. There is also backscattering of
particles from the solenoid coil at large radius back into the 
TPC, however, tracks from backscattering are eliminated by the 
requirement that they originate from the target.

Other than for the transverse momentum $p_{\rm T}$, migration 
is nearly negligible in the measurement of the
polar angle $\theta$. 

Charge misidentification occurs only 
at large transverse momentum, at the
level of a few per cent. For example, 
a few `antiprotons' at large transverse
momentum are charge-misidentified protons and 
treated accordingly in the migration correction.

Pion decay into muons occurs at the typical level of 2\%. 
When the pion decay occurs in the first few centimetres of the 
flight path, the phenomenon is taken care
of by the migration correction. When the pion decay occurs later,
the track is likely to be lost because of the requirement
that it originates from the target. Therefore, each pion receives a
weight that compensates on the average for the loss from decay 
along a path of 200~mm length.

The re-interaction of secondaries takes place in the
target material or in other materials between the target and
the TPC volume. The typical probability for re-interaction
is 3\% for the former, and 2\% for the latter. The
re-interaction leads to tracks with other parameters
than the initial track, and is taken care of by the
migration correction.

\subsection{Systematic errors}

The systematic precision of our inclusive cross-sections 
is at the few-per-cent level, from errors
in the normalization, in the momentum measurement, in
particle identification, and in the corrections applied
to the data.

The systematic error of the absolute flux normalization is 
taken as 2\%. This error arises from uncertainties in the
target thickness, in the contribution of large-angle 
scattering of beam particles, in the attenuation of beam particles in the target, and in the subtraction of
the muon and electron contaminations. Another contribution comes from the removal of events with an abnormally large number of TPC hits above threshold.

The systematic error of the track finding  
efficiency is taken as 1\% which reflects differences 
between results from different persons who conducted
eyeball scans. We also take the statistical errors of
the parameters of a fit to scan results as shown in 
Fig.~\ref{reconstructionefficiency} (a)  
as systematic error into account.
The systematic error of the correction 
for losses from the requirement of at least 10 TPC clusters 
per track is taken as 20\% of the correction which 
itself is in the range of 5 to 30\%. This estimate arose
from differences between the four TPC sectors that
were used in our analysis, and from the observed 
variations with time. 

The systematic error of the $p_{\rm T}$ scale is taken as
2\% as discussed in Ref.~\cite{TPCpub}.

The systematic errors of the proton, pion, and electron
abundances are taken as 10\%. We stress that errors on 
abundances only lead to cross-section errors in case of a strong overlap of the resolution functions
of both identification variables, \dedx\ and $\beta$. 
The systematic error of the correction for migration, absorption
of secondary protons and pions in materials, and for pion
decay into muons, is taken as 20\% of the correction, or 1\% of the cross-section, whichever is larger. These estimates reflect our experience 
with remanent differences between data and Monte Carlo 
simulations after weighting Monte Carlo events with smooth functions 
with a view to reproducing the data simultaneously in 
several variables in the best possible way.

All systematic errors are propagated into the momentum 
spectra of secondaries and then added in quadrature.

\section{Kaon and deuteron production}
\label{kaonsanddeuterons}

The statistics from the $+8.9$~GeV/{\it c} beam on a 
5\% $\lambda_{\rm abs}$ beryllium target is much larger than
for any other combination of beam and target. This permits
us to investigate in this particular data set the production
of K$^+$'s and deuterons in addition to the
dominant protons, $\pi^+$'s, and $\pi^-$'s. 
With a view to benefiting from the cancellation of systematic 
errors, we present results in terms of the ratios K$^+$/$\pi^+$ and 
d/p.  

\subsection{Kaons}

\begin{figure}[htp]
\begin{center}
\includegraphics[width=0.5\textwidth]{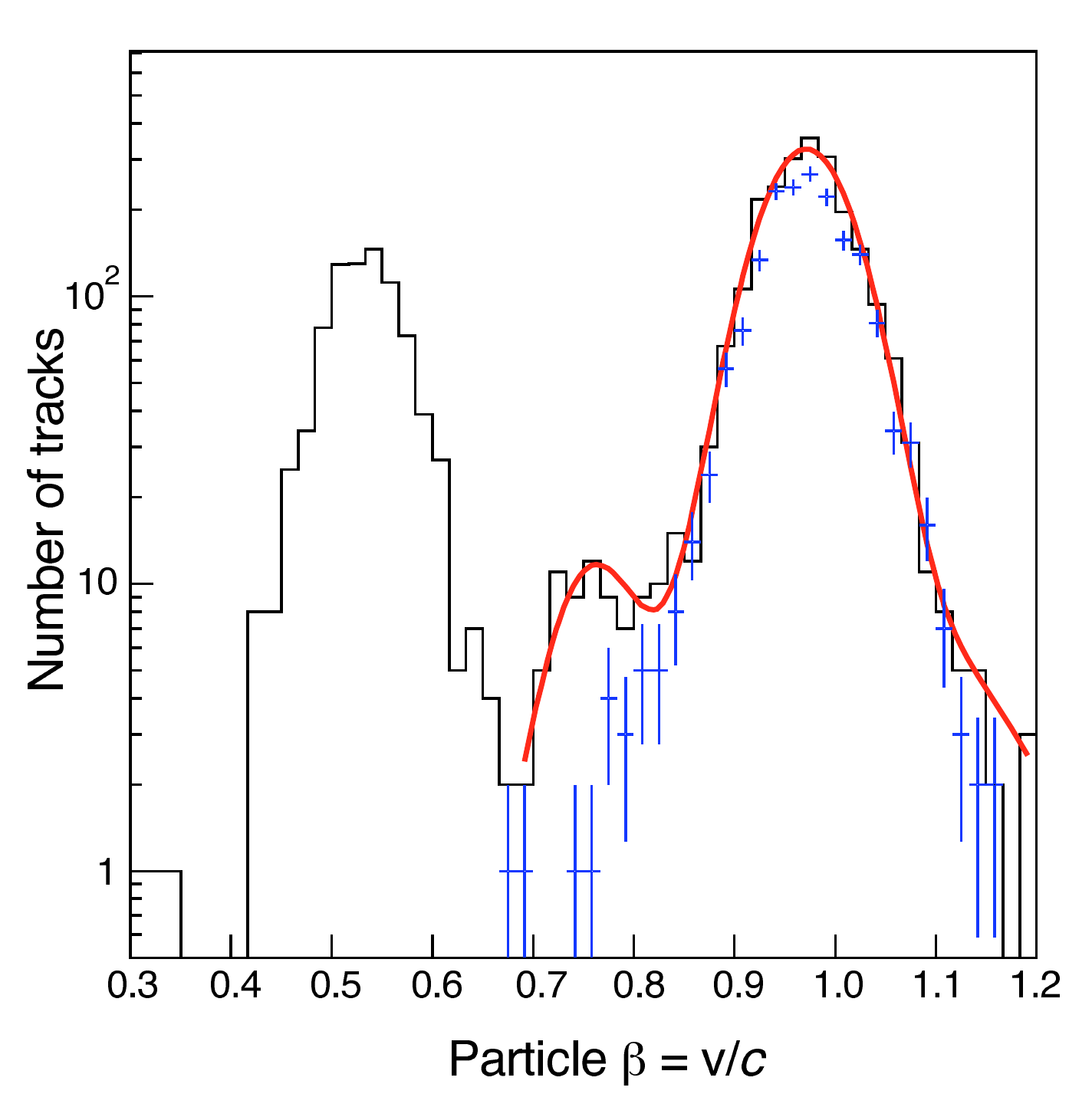} 
\caption{Distribution of $\beta$ of positive secondaries on a logarithmic scale, with the K$^+$ signal showing up between pions and protons; the crosses show the $\beta$ distribution of $\pi^-$'s; the shown fit is explained in the text.}
\label{betaofkaons}
\end{center}
\end{figure}

Figure~\ref{betaofkaons} shows the relative velocity $\beta$ 
of positive secondaries for the
polar-angle range $20.5^\circ < \theta < 25.3^\circ$
and momentum between 520 and 560~MeV/{\it c}.
A logarithmic scale is employed to 
make K$^+$ production visible
which is at the level of a few per cent 
of the $\pi^+$ production.  
The K$^+$ signal shows up between the proton and $\pi^+$
signal thanks to the good resolution of
the $\beta$ measurement by the RPCs. 
The K$^+$ signal is fitted with a Gaussian. The
$\pi^+$ signal is represented by a Gaussian 
together with a tail that is experimentally determined
from the $\beta$ distribution of the $\pi^-$'s. The latter
is shown with crosses in Fig.~\ref{betaofkaons}. A possible 
K$^-$ contribution is minimized by a \dedx\ cut.

In order to maximize the time of flight and hence the separation
power, we restrict the analysis to 
the forward region in the
range $20^\circ < \theta < 32^\circ$. The momentum
is required to be in the range $400 < p < 700$~MeV/{\it c},
and \dedx\ must be between 70\% and 155\% of the nominal
value. 

Several corrections must be made to the fit results of the
relative K$^+$ abundance. Correcting for cuts on the charge
of the RPC signal,
made with a view to optimizing time-of-flight resolution, 
reduces the signal by 5\%. The correction for 
the non-Gaussian tail of the $\beta$ distribution of K$^+$'s 
increases the signal by
8\%. The correction for different absorption of K$^+$'s and $\pi^+$'s
in structural materials increases the signal by 1\%.

\begin{figure}[ht]
\begin{center}
\includegraphics[width=0.5\textwidth]{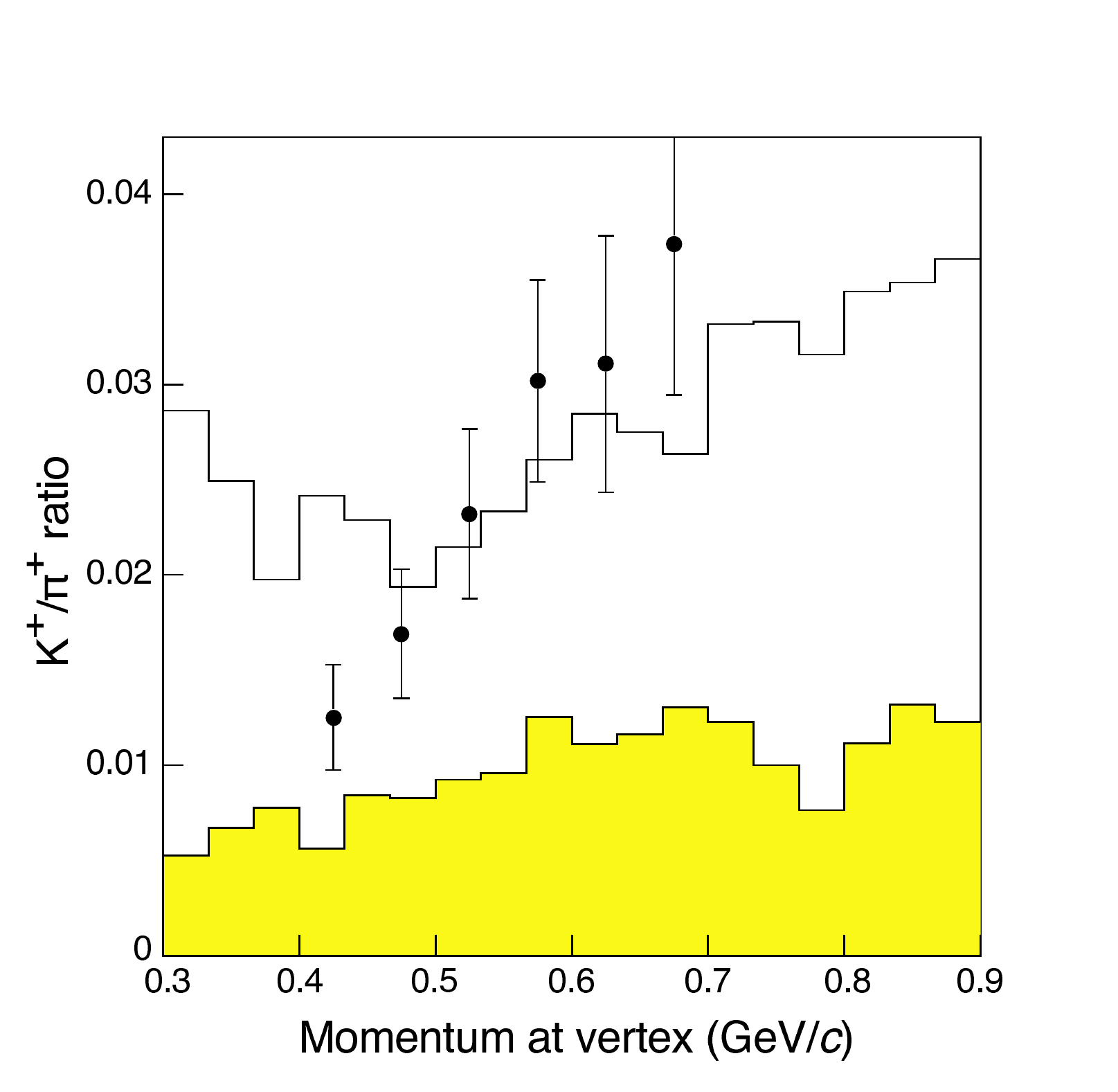} 
\caption{Ratio of K$^+$/$\pi^+$ in $+8.9$~GeV/{\it c} proton and
$\pi^+$ interactions with beryllium nuclei; the data points
(black circles) are closer to the prediction by
the FRITIOF model (open histogram) in Geant4, 
but agree poorly with its Binary Cascade model (shaded histogram).}
\label{kaonstopions}
\end{center}
\end{figure}

Altogether, the resulting ratio is      
\begin{displaymath}
R_{\rm K} (p, \theta) = \frac
{{\rm d}^2 \sigma_{\rm K} (p,\theta) / {\rm d} p {\rm d} \Omega} 
{{\rm d}^2 \sigma_\pi (p,\theta) / {\rm d} p {\rm d} \Omega} = 
0.020 \pm 0.003   \; ,
\end{displaymath}
averaged over the said range of momentum and polar angle,
and over the proton and $\pi^+$ beams (in kaon production,
no significant difference is seen between these beams).
Figure~\ref{kaonstopions} shows the K$^+$/$\pi^+$ ratio as 
a function of particle momentum and compares the measured
ratios with the ratios from the FRITIOF and 
Binary Cascade hadron production models in Geant4. The data points
are closer to the prediction by the
FRITIOF model, however, the dependence on momentum is not reproduced. 
The agreement with the Binary Cascade model is poor. 

\subsection{Deuterons}

\begin{figure}[ht]
\begin{center}
\includegraphics[width=0.5\textwidth]{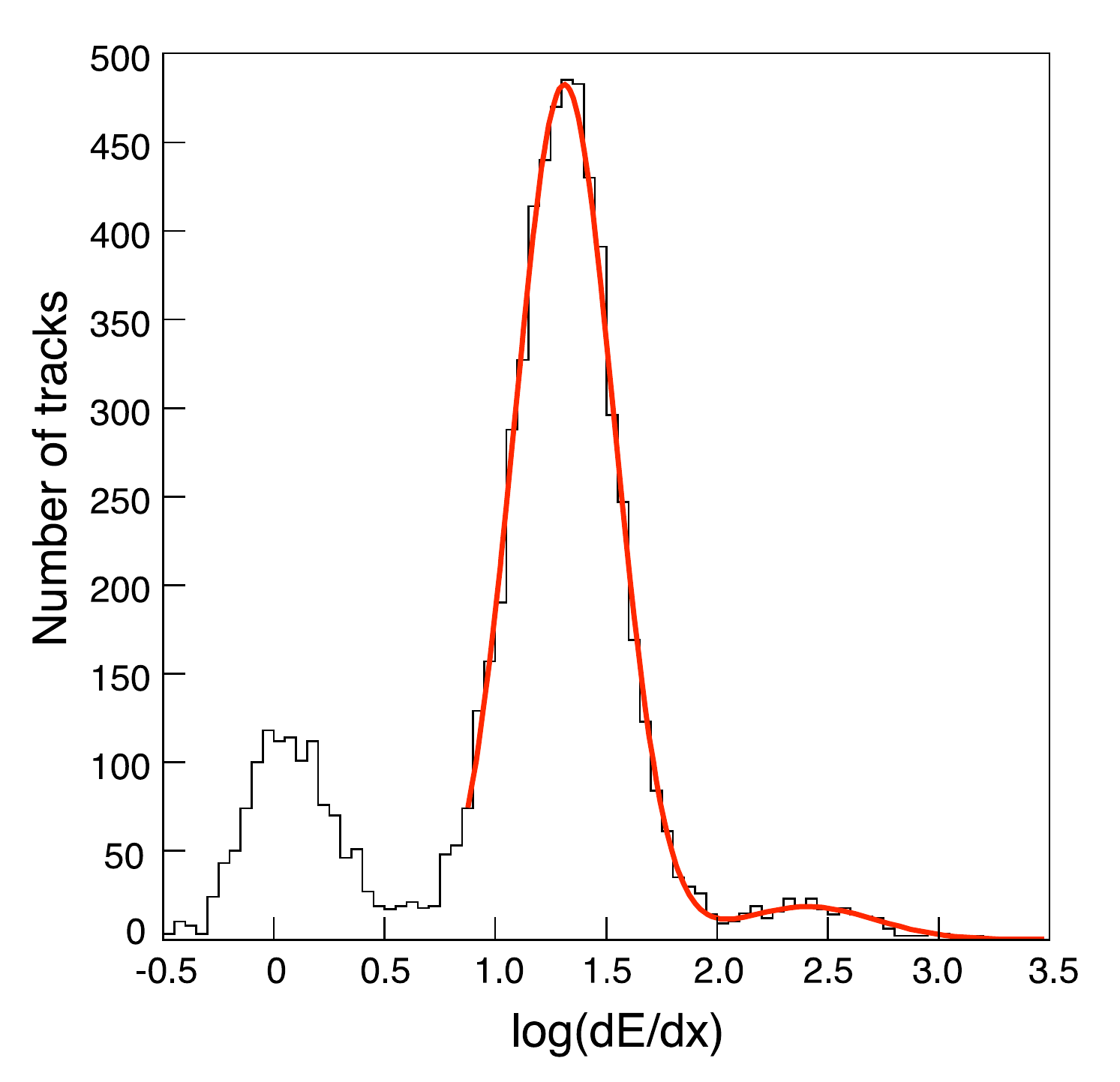} 
\caption{Distribution of \dedx\ of positive secondaries, with the deuteron signal showing up at large \dedx ;
see the text for the cuts applied; both the proton and deuteron signals 
are fitted with a Gaussian.}
\label{fitdEdxfordeuterons}
\end{center}
\end{figure}

Figure~\ref{fitdEdxfordeuterons} shows the \dedx\ 
of positive secondaries for the polar-angle range
$30^\circ < \theta < 45^\circ$ and the momentum
range from 500 to 600~MeV/{\it c} (this momentum
range refers to the momentum measured in the TPC and not to
the momentum at the vertex). Pions and electrons
are reduced by a loose time-of-flight cut. A clear signal
of deuterons is visible at large \dedx , next to the 
abundant protons. 

In order to transform the ratio measured in the TPC 
volume to that at the vertex, appropriate corrections 
for the different energy loss of protons and deuterons in materials 
between the vertex and the TPC volume, and for differences
in the momentum spectra of protons and deuterons, must 
be applied. The results for the
ratio      
\begin{displaymath}
R_{\rm d} (p, \theta) = \frac
{{\rm d}^2 \sigma_{\rm d} (p,\theta) / {\rm d} p {\rm d} \Omega} 
{{\rm d}^2 \sigma_{\rm p} (p,\theta) / {\rm d} p {\rm d} \Omega}   \; ,
\end{displaymath}
averaged over the momentum at the vertex between 600~MeV/{\it c} 
and 1050~MeV/{\it c}, are given in 
Table~\ref{dtopratio}.
\begin{table}[h]
\caption{Ratio $R_{\rm d}$ of deuterons to protons, for different
beam particles, averaged over the momentum at the vertex 
between 600~MeV/{\it c} and 1050~MeV/{\it c}.}
\label{dtopratio}
\begin{center}
\begin{tabular}{|l|c|c|c|}
\hline
     & $+8.9$~GeV/{\it c} protons & $+8.9$~GeV/{\it c} $\pi^+$ 
     & $-8.0$~GeV/{\it c} $\pi^-$  \\ 
\hline    
\hline
$20^\circ<\theta<30^\circ$  & $0.051 \pm 0.004$ & $0.038 \pm 0.004$ 
                            & $0.05 \pm 0.01$   \\   
$30^\circ<\theta<45^\circ$  & $0.076 \pm 0.004$ & $0.065 \pm 0.005$    
                            & $0.07 \pm 0.01$   \\
$45^\circ<\theta<65^\circ$  & $0.113 \pm 0.009$ & $0.094 \pm 0.008$  
                            & $0.13 \pm 0.02$   \\
$65^\circ<\theta<90^\circ$  & $0.17 \pm 0.03$   & $0.15 \pm 0.03$    
                            & $0.16 \pm 0.03$   \\
$90^\circ<\theta<125^\circ$ & $0.26 \pm 0.04$   & $0.21 \pm 0.05$    
                            & $0.30 \pm 0.05$   \\
\hline
\end{tabular}
\end{center}
\end{table}

\begin{figure}[htp]
\begin{center}
\includegraphics[width=0.5\textwidth]{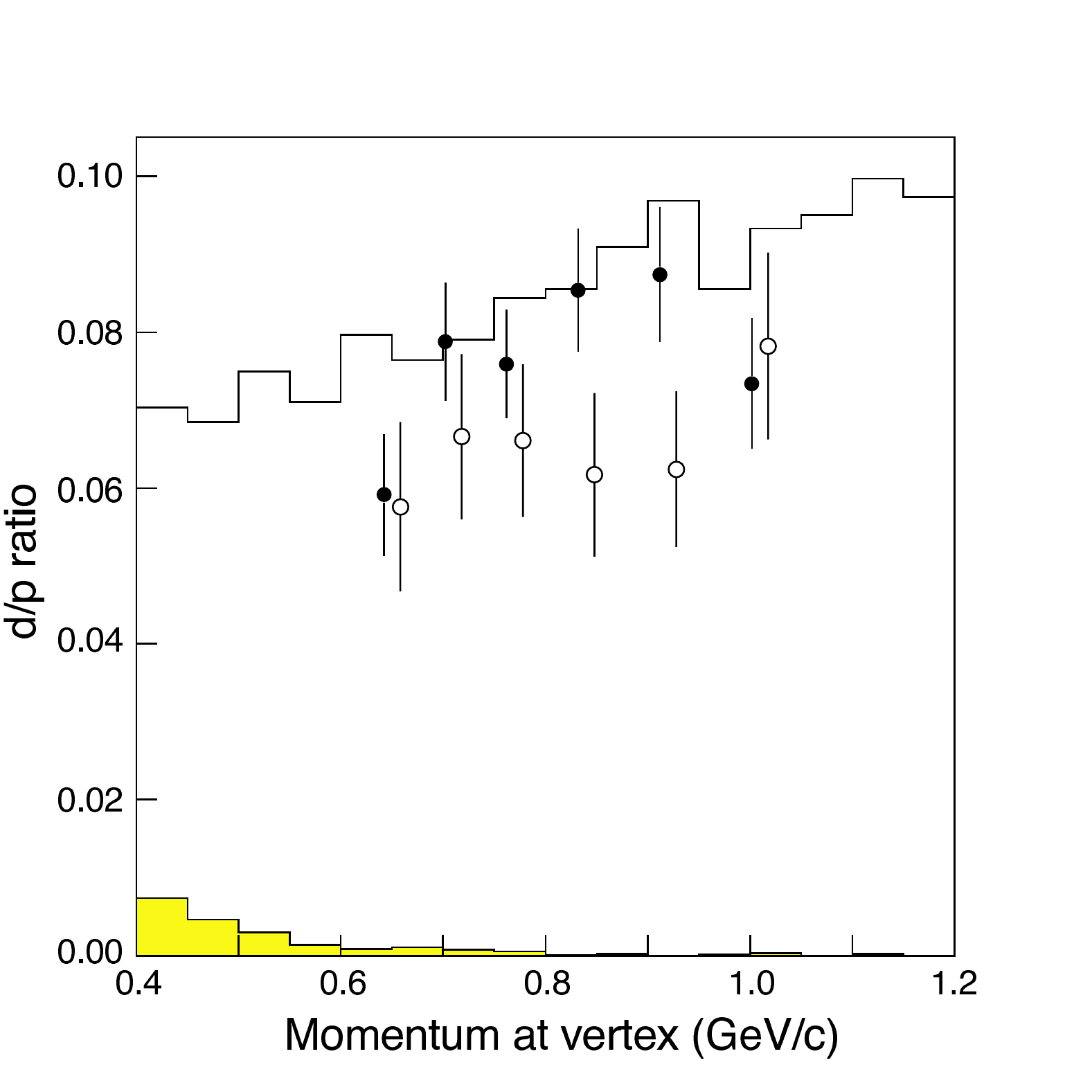} 
\caption{Ratio $R_{\rm d}$ in $+8.9$~GeV/{\it c} proton and
$\pi^+$ interactions with beryllium nuclei, for 
$30^\circ < \theta < 45^\circ$, as a 
function of the momentum at the vertex; black data points
refer to the proton beam, open circles to the $\pi^+$ beam.
The proton beam data are compared with the predictions of the 
FRITIOF (open histogram) and the Binary Cascade (shaded histogram)
models in Geant4.}
\label{deuteronstopions}
\end{center}
\end{figure}

The ratios $R_{\rm d}$ for the $+8.9$~GeV/{\it c} proton 
and $\pi^+$ beams are shown in Fig.~\ref{deuteronstopions} 
for the polar-angle range $30^\circ < \theta < 45^\circ$ as a 
function of the momentum at the vertex. 
We note that the deuteron abundance is resonably well 
reproduced by the FRITIOF String Fragmentation model 
used in the Geant4 simulation tool kit, while it is
underestimated by about one order 
of magnitude by the Binary Cascade model.
%


\section{Double-differential inclusive cross-sections of protons and pions}
\label{cross-sections}

In Tables~\ref{pro.probe8}--\ref{pim.pimbe8} we give
the double-differential inclusive cross-sections 
${\rm d}^2 \sigma / {\rm d} p {\rm d} \Omega$
for all nine combinations of
incoming beam particle and secondary particle, including
statistical and systematic errors. In each bin,  
the average momentum and the average polar angle are also given.

Cross-sections are only given if the total error is not larger than the cross-section itself.
Since our track reconstruction algorithm is optimized for
tracks with $p_{\rm T}$ above $\sim$70~MeV/{\it c} in the
TPC volume, we do not give cross-sections from tracks with $p_{\rm T}$ 
below this value.
Because of the absorption of slow protons in the material between the
vertex and the TPC gas, and with a view to keeping the correction
for absorption losses below 30\%, cross-sections from protons are 
limited to $p > 350$~MeV/{\it c} at the interaction vertex. 
Proton cross-sections are also not given if a 10\% error on the proton energy loss in materials between the interaction vertex and the TPC volume leads to a momentum change larger than 2\%.
Pion cross-sections are not given if pions are separated from protons by less than twice the time-of-flight resolution.

The data given in Tables~\ref{pro.probe8}--\ref{pim.pimbe8} are available
in ASCII format in Ref.~\cite{ASCIIdatatables}.

 \begin{table}[h]
 \begin{scriptsize}
 \caption{Double-differential inclusive
  cross-section ${\rm d}^2 \sigma /{\rm d}p{\rm d}\Omega$
  [mb/(GeV/{\it c} sr)] of the production of protons
  in p + Be $\rightarrow$ p + X interactions
  with $+8.9$~GeV/{\it c} beam momentum;
  the first error is statistical, the second systematic; 
 $p_{\rm T}$ in GeV/{\it c}, polar angle $\theta$ in degrees.}
 \label{pro.probe8}
 \begin{center}
\begin{tabular}{|c||c|c|rcrcr||c|c|rcrcr|}
 \hline
   & \multicolumn{7}{c||}{$20<\theta<30$}
  & \multicolumn{7}{c|}{$30<\theta<40$} \\
 \hline
 $p_{\rm T}$ & $\langle p_{\rm T} \rangle$ & $\langle \theta \rangle$
  & \multicolumn{5}{c||}{${\rm d}^2 \sigma /{\rm d}p{\rm d}\Omega$}
  &$\langle p_{\rm T} \rangle$ & $\langle \theta \rangle$
  & \multicolumn{5}{c|}{${\rm d}^2 \sigma /{\rm d}p{\rm d}\Omega$} \\
 \hline
0.20--0.24 & 0.220 &  24.9 &  48.38 & $\!\!\pm\!\!$ &  0.72 & $\!\!\pm\!\!$ &  2.25 &  &  &  \multicolumn{5}{c|}{ } \\
0.24--0.30 & 0.270 &  24.9 &  48.68 & $\!\!\pm\!\!$ &  0.59 & $\!\!\pm\!\!$ &  2.02 & 0.271 &  35.0 &  42.76 & $\!\!\pm\!\!$ &  0.54 & $\!\!\pm\!\!$ &  1.65 \\
0.30--0.36 & 0.330 &  25.0 &  44.77 & $\!\!\pm\!\!$ &  0.55 & $\!\!\pm\!\!$ &  1.66 & 0.329 &  35.0 &  41.00 & $\!\!\pm\!\!$ &  0.52 & $\!\!\pm\!\!$ &  1.39 \\
0.36--0.42 & 0.389 &  24.9 &  41.06 & $\!\!\pm\!\!$ &  0.53 & $\!\!\pm\!\!$ &  1.41 & 0.389 &  35.0 &  38.10 & $\!\!\pm\!\!$ &  0.51 & $\!\!\pm\!\!$ &  1.20 \\
0.42--0.50 & 0.459 &  24.9 &  39.55 & $\!\!\pm\!\!$ &  0.45 & $\!\!\pm\!\!$ &  1.26 & 0.459 &  34.9 &  33.39 & $\!\!\pm\!\!$ &  0.42 & $\!\!\pm\!\!$ &  1.00 \\
0.50--0.60 & 0.548 &  24.9 &  34.14 & $\!\!\pm\!\!$ &  0.37 & $\!\!\pm\!\!$ &  1.11 & 0.548 &  34.9 &  27.89 & $\!\!\pm\!\!$ &  0.34 & $\!\!\pm\!\!$ &  0.87 \\
\hline
 \end{tabular}
 \end{center}
 \end{scriptsize}
 \end{table}

 \begin{table}[h]
 \begin{scriptsize}
 \caption{Double-differential inclusive
  cross-section ${\rm d}^2 \sigma /{\rm d}p{\rm d}\Omega$
  [mb/(GeV/{\it c} sr)] of the production of $\pi^+$'s
  in p + Be $\rightarrow$ $\pi^+$ + X interactions
  with $+8.9$~GeV/{\it c} beam momentum;
  the first error is statistical, the second systematic; 
 $p_{\rm T}$ in GeV/{\it c}, polar angle $\theta$ in degrees.}
 \label{pip.probe8}
 \begin{center}
 \begin{tabular}{|c||c|c|rcrcr||c|c|rcrcr|}
 \hline
   & \multicolumn{7}{c||}{$20<\theta<30$}
  & \multicolumn{7}{c|}{$30<\theta<40$} \\
 \hline
 $p_{\rm T}$ & $\langle p_{\rm T} \rangle$ & $\langle \theta \rangle$
  & \multicolumn{5}{c||}{${\rm d}^2 \sigma /{\rm d}p{\rm d}\Omega$}
  &$\langle p_{\rm T} \rangle$ & $\langle \theta \rangle$
  & \multicolumn{5}{c|}{${\rm d}^2 \sigma /{\rm d}p{\rm d}\Omega$} \\
 \hline
0.10--0.13 & 0.116 &  24.8 &  53.94 & $\!\!\pm\!\!$ &  1.04 & $\!\!\pm\!\!$ &  3.98 & 0.116 &  34.7 &  37.12 & $\!\!\pm\!\!$ &  0.84 & $\!\!\pm\!\!$ &  2.54 \\
0.13--0.16 & 0.146 &  24.5 &  68.32 & $\!\!\pm\!\!$ &  1.08 & $\!\!\pm\!\!$ &  3.91 & 0.146 &  34.7 &  48.46 & $\!\!\pm\!\!$ &  0.92 & $\!\!\pm\!\!$ &  2.53 \\
0.16--0.20 & 0.180 &  24.7 &  80.65 & $\!\!\pm\!\!$ &  0.98 & $\!\!\pm\!\!$ &  3.65 & 0.180 &  34.8 &  51.60 & $\!\!\pm\!\!$ &  0.78 & $\!\!\pm\!\!$ &  2.26 \\
0.20--0.24 & 0.220 &  24.7 &  83.11 & $\!\!\pm\!\!$ &  0.98 & $\!\!\pm\!\!$ &  3.08 & 0.220 &  34.7 &  55.60 & $\!\!\pm\!\!$ &  0.80 & $\!\!\pm\!\!$ &  2.10 \\
0.24--0.30 & 0.269 &  24.7 &  76.59 & $\!\!\pm\!\!$ &  0.76 & $\!\!\pm\!\!$ &  2.44 & 0.269 &  34.7 &  51.20 & $\!\!\pm\!\!$ &  0.61 & $\!\!\pm\!\!$ &  1.60 \\
0.30--0.36 & 0.329 &  24.7 &  63.92 & $\!\!\pm\!\!$ &  0.69 & $\!\!\pm\!\!$ &  1.83 & 0.329 &  34.7 &  43.01 & $\!\!\pm\!\!$ &  0.56 & $\!\!\pm\!\!$ &  1.19 \\
\hline
 \end{tabular}
 \end{center}
 \end{scriptsize}
 \end{table}

 \begin{table}[h]
 \begin{scriptsize}
 \caption{Double-differential inclusive
  cross-section ${\rm d}^2 \sigma /{\rm d}p{\rm d}\Omega$
  [mb/(GeV/{\it c} sr)] of the production of $\pi^-$'s
  in p + Be $\rightarrow$ $\pi^-$ + X interactions
  with $+8.9$~GeV/{\it c} beam momentum;
  the first error is statistical, the second systematic; 
 $p_{\rm T}$ in GeV/{\it c}, polar angle $\theta$ in degrees.}
 \label{pim.probe8}
 \begin{center}
 \begin{tabular}{|c||c|c|rcrcr||c|c|rcrcr|}
 \hline
   & \multicolumn{7}{c||}{$20<\theta<30$}
  & \multicolumn{7}{c|}{$30<\theta<40$} \\
 \hline
 $p_{\rm T}$ & $\langle p_{\rm T} \rangle$ & $\langle \theta \rangle$
  & \multicolumn{5}{c||}{${\rm d}^2 \sigma /{\rm d}p{\rm d}\Omega$}
  &$\langle p_{\rm T} \rangle$ & $\langle \theta \rangle$
  & \multicolumn{5}{c|}{${\rm d}^2 \sigma /{\rm d}p{\rm d}\Omega$} \\
 \hline
0.10--0.13 & 0.116 &  24.7 &  52.88 & $\!\!\pm\!\!$ &  0.99 & $\!\!\pm\!\!$ &  3.67 & 0.116 &  34.7 &  37.22 & $\!\!\pm\!\!$ &  0.80 & $\!\!\pm\!\!$ &  2.54 \\
0.13--0.16 & 0.145 &  24.8 &  68.86 & $\!\!\pm\!\!$ &  1.07 & $\!\!\pm\!\!$ &  3.72 & 0.145 &  34.7 &  43.73 & $\!\!\pm\!\!$ &  0.81 & $\!\!\pm\!\!$ &  2.33 \\
0.16--0.20 & 0.180 &  24.7 &  73.99 & $\!\!\pm\!\!$ &  0.92 & $\!\!\pm\!\!$ &  3.23 & 0.180 &  34.7 &  51.74 & $\!\!\pm\!\!$ &  0.77 & $\!\!\pm\!\!$ &  2.28 \\
0.20--0.24 & 0.220 &  24.8 &  70.71 & $\!\!\pm\!\!$ &  0.89 & $\!\!\pm\!\!$ &  2.59 & 0.220 &  34.6 &  50.06 & $\!\!\pm\!\!$ &  0.74 & $\!\!\pm\!\!$ &  1.85 \\
0.24--0.30 & 0.269 &  24.8 &  63.44 & $\!\!\pm\!\!$ &  0.68 & $\!\!\pm\!\!$ &  1.95 & 0.269 &  34.7 &  45.71 & $\!\!\pm\!\!$ &  0.57 & $\!\!\pm\!\!$ &  1.41 \\
0.30--0.36 & 0.329 &  24.8 &  50.19 & $\!\!\pm\!\!$ &  0.60 & $\!\!\pm\!\!$ &  1.37 & 0.329 &  34.8 &  38.12 & $\!\!\pm\!\!$ &  0.52 & $\!\!\pm\!\!$ &  1.04 \\
\hline
 \end{tabular}
 \end{center}
 \end{scriptsize}
 \end{table}

 \begin{table}[h]
 \begin{scriptsize}
 \caption{Double-differential inclusive
  cross-section ${\rm d}^2 \sigma /{\rm d}p{\rm d}\Omega$
  [mb/(GeV/{\it c} sr)] of the production of protons
  in $\pi^+$ + Be $\rightarrow$ p + X interactions
  with $+8.9$~GeV/{\it c} beam momentum;
  the first error is statistical, the second systematic; 
 $p_{\rm T}$ in GeV/{\it c}, polar angle $\theta$ in degrees.}
 \label{pro.pipbe8}
 \begin{center}
 \begin{tabular}{|c||c|c|rcrcr||c|c|rcrcr|}
 \hline
   & \multicolumn{7}{c||}{$20<\theta<30$}
  & \multicolumn{7}{c|}{$30<\theta<40$} \\
 \hline
 $p_{\rm T}$ & $\langle p_{\rm T} \rangle$ & $\langle \theta \rangle$
  & \multicolumn{5}{c||}{${\rm d}^2 \sigma /{\rm d}p{\rm d}\Omega$}
  &$\langle p_{\rm T} \rangle$ & $\langle \theta \rangle$
  & \multicolumn{5}{c|}{${\rm d}^2 \sigma /{\rm d}p{\rm d}\Omega$} \\
 \hline
0.20--0.24 & 0.220 &  24.8 &  34.44 & $\!\!\pm\!\!$ &  0.76 & $\!\!\pm\!\!$ &  1.62 &  &  &  \multicolumn{5}{c|}{ } \\
0.24--0.30 & 0.269 &  25.1 &  36.77 & $\!\!\pm\!\!$ &  0.64 & $\!\!\pm\!\!$ &  1.54 & 0.270 &  34.9 &  32.90 & $\!\!\pm\!\!$ &  0.66 & $\!\!\pm\!\!$ &  1.30 \\
0.30--0.36 & 0.329 &  25.0 &  32.47 & $\!\!\pm\!\!$ &  0.58 & $\!\!\pm\!\!$ &  1.22 & 0.330 &  34.9 &  30.81 & $\!\!\pm\!\!$ &  0.56 & $\!\!\pm\!\!$ &  1.07 \\
0.36--0.42 & 0.389 &  24.9 &  30.62 & $\!\!\pm\!\!$ &  0.56 & $\!\!\pm\!\!$ &  1.05 & 0.389 &  34.9 &  27.98 & $\!\!\pm\!\!$ &  0.54 & $\!\!\pm\!\!$ &  0.89 \\
0.42--0.50 & 0.458 &  25.0 &  28.19 & $\!\!\pm\!\!$ &  0.47 & $\!\!\pm\!\!$ &  0.90 & 0.459 &  34.9 &  25.72 & $\!\!\pm\!\!$ &  0.45 & $\!\!\pm\!\!$ &  0.78 \\
0.50--0.60 & 0.547 &  24.9 &  23.47 & $\!\!\pm\!\!$ &  0.37 & $\!\!\pm\!\!$ &  0.77 & 0.548 &  34.9 &  20.87 & $\!\!\pm\!\!$ &  0.36 & $\!\!\pm\!\!$ &  0.65 \\
\hline
 \end{tabular}
 \end{center}
 \end{scriptsize}
 \end{table}

 \begin{table}[h]
 \begin{scriptsize}
 \caption{Double-differential inclusive
  cross-section ${\rm d}^2 \sigma /{\rm d}p{\rm d}\Omega$
  [mb/(GeV/{\it c} sr)] of the production of $\pi^+$'s
  in $\pi^+$ + Be $\rightarrow$ $\pi^+$ + X interactions
  with $+8.9$~GeV/{\it c} beam momentum;
  the first error is statistical, the second systematic; 
 $p_{\rm T}$ in GeV/{\it c}, polar angle $\theta$ in degrees.}
 \label{pip.pipbe8}
 \begin{center}
 \begin{tabular}{|c||c|c|rcrcr||c|c|rcrcr|}
 \hline
   & \multicolumn{7}{c||}{$20<\theta<30$}
  & \multicolumn{7}{c|}{$30<\theta<40$} \\
 \hline
 $p_{\rm T}$ & $\langle p_{\rm T} \rangle$ & $\langle \theta \rangle$
  & \multicolumn{5}{c||}{${\rm d}^2 \sigma /{\rm d}p{\rm d}\Omega$}
  &$\langle p_{\rm T} \rangle$ & $\langle \theta \rangle$
  & \multicolumn{5}{c|}{${\rm d}^2 \sigma /{\rm d}p{\rm d}\Omega$} \\
 \hline
0.10--0.13 & 0.115 &  24.6 &  52.73 & $\!\!\pm\!\!$ &  1.28 & $\!\!\pm\!\!$ &  3.46 & 0.116 &  34.7 &  38.43 & $\!\!\pm\!\!$ &  1.06 & $\!\!\pm\!\!$ &  2.58 \\
0.13--0.16 & 0.145 &  24.6 &  68.39 & $\!\!\pm\!\!$ &  1.35 & $\!\!\pm\!\!$ &  3.63 & 0.146 &  34.7 &  49.02 & $\!\!\pm\!\!$ &  1.16 & $\!\!\pm\!\!$ &  2.55 \\
0.16--0.20 & 0.180 &  24.6 &  85.35 & $\!\!\pm\!\!$ &  1.26 & $\!\!\pm\!\!$ &  3.81 & 0.180 &  34.7 &  54.25 & $\!\!\pm\!\!$ &  0.99 & $\!\!\pm\!\!$ &  2.36 \\
0.20--0.24 & 0.220 &  24.7 &  90.83 & $\!\!\pm\!\!$ &  1.28 & $\!\!\pm\!\!$ &  3.47 & 0.220 &  34.7 &  59.18 & $\!\!\pm\!\!$ &  1.03 & $\!\!\pm\!\!$ &  2.21 \\
0.24--0.30 & 0.270 &  24.6 &  86.94 & $\!\!\pm\!\!$ &  1.01 & $\!\!\pm\!\!$ &  2.86 & 0.270 &  34.7 &  55.39 & $\!\!\pm\!\!$ &  0.80 & $\!\!\pm\!\!$ &  1.77 \\
0.30--0.36 & 0.329 &  24.7 &  78.97 & $\!\!\pm\!\!$ &  0.96 & $\!\!\pm\!\!$ &  2.38 & 0.329 &  34.6 &  49.51 & $\!\!\pm\!\!$ &  0.75 & $\!\!\pm\!\!$ &  1.41 \\
\hline
 \end{tabular}
 \end{center}
 \end{scriptsize}
 \end{table}

 \begin{table}[h]
 \begin{scriptsize}
 \caption{Double-differential inclusive
  cross-section ${\rm d}^2 \sigma /{\rm d}p{\rm d}\Omega$
  [mb/(GeV/{\it c} sr)] of the production of $\pi^-$'s
  in $\pi^+$ + Be $\rightarrow$ $\pi^-$ + X interactions
  with $+8.9$~GeV/{\it c} beam momentum;
  the first error is statistical, the second systematic; 
 $p_{\rm T}$ in GeV/{\it c}, polar angle $\theta$ in degrees.}
 \label{pim.pipbe8}
 \begin{center}
 \begin{tabular}{|c||c|c|rcrcr||c|c|rcrcr|}
 \hline
   & \multicolumn{7}{c||}{$20<\theta<30$}
  & \multicolumn{7}{c|}{$30<\theta<40$} \\
 \hline
 $p_{\rm T}$ & $\langle p_{\rm T} \rangle$ & $\langle \theta \rangle$
  & \multicolumn{5}{c||}{${\rm d}^2 \sigma /{\rm d}p{\rm d}\Omega$}
  &$\langle p_{\rm T} \rangle$ & $\langle \theta \rangle$
  & \multicolumn{5}{c|}{${\rm d}^2 \sigma /{\rm d}p{\rm d}\Omega$} \\
 \hline
0.10--0.13 & 0.116 &  24.8 &  46.25 & $\!\!\pm\!\!$ &  1.15 & $\!\!\pm\!\!$ &  3.10 & 0.116 &  34.8 &  33.42 & $\!\!\pm\!\!$ &  0.94 & $\!\!\pm\!\!$ &  2.33 \\
0.13--0.16 & 0.146 &  24.6 &  61.24 & $\!\!\pm\!\!$ &  1.26 & $\!\!\pm\!\!$ &  3.17 & 0.145 &  34.6 &  36.42 & $\!\!\pm\!\!$ &  0.92 & $\!\!\pm\!\!$ &  1.93 \\
0.16--0.20 & 0.180 &  24.7 &  65.33 & $\!\!\pm\!\!$ &  1.07 & $\!\!\pm\!\!$ &  2.89 & 0.180 &  34.6 &  43.91 & $\!\!\pm\!\!$ &  0.88 & $\!\!\pm\!\!$ &  1.95 \\
0.20--0.24 & 0.220 &  24.6 &  70.02 & $\!\!\pm\!\!$ &  1.10 & $\!\!\pm\!\!$ &  2.61 & 0.220 &  34.6 &  44.85 & $\!\!\pm\!\!$ &  0.87 & $\!\!\pm\!\!$ &  1.69 \\
0.24--0.30 & 0.269 &  24.7 &  63.57 & $\!\!\pm\!\!$ &  0.85 & $\!\!\pm\!\!$ &  2.03 & 0.270 &  34.7 &  42.49 & $\!\!\pm\!\!$ &  0.69 & $\!\!\pm\!\!$ &  1.35 \\
0.30--0.36 & 0.329 &  24.7 &  51.65 & $\!\!\pm\!\!$ &  0.76 & $\!\!\pm\!\!$ &  1.45 & 0.329 &  34.7 &  35.06 & $\!\!\pm\!\!$ &  0.62 & $\!\!\pm\!\!$ &  0.99 \\
\hline
 \end{tabular}
 \end{center}
 \end{scriptsize}
 \end{table}

 \begin{table}[h]
 \begin{scriptsize}
 \caption{Double-differential inclusive
  cross-section ${\rm d}^2 \sigma /{\rm d}p{\rm d}\Omega$
  [mb/(GeV/{\it c} sr)] of the production of protons
  in $\pi^-$ + Be $\rightarrow$ p + X interactions
  with $-8.0$~GeV/{\it c} beam momentum;
  the first error is statistical, the second systematic; 
 $p_{\rm T}$ in GeV/{\it c}, polar angle $\theta$ in degrees.}
 \label{pro.pimbe8}
 \begin{center}
 \begin{tabular}{|c||c|c|rcrcr||c|c|rcrcr|}
 \hline
   & \multicolumn{7}{c||}{$20<\theta<30$}
  & \multicolumn{7}{c|}{$30<\theta<40$} \\
 \hline
 $p_{\rm T}$ & $\langle p_{\rm T} \rangle$ & $\langle \theta \rangle$
  & \multicolumn{5}{c||}{${\rm d}^2 \sigma /{\rm d}p{\rm d}\Omega$}
  &$\langle p_{\rm T} \rangle$ & $\langle \theta \rangle$
  & \multicolumn{5}{c|}{${\rm d}^2 \sigma /{\rm d}p{\rm d}\Omega$} \\
 \hline
0.20--0.24 & 0.221 &  25.1 &  28.46 & $\!\!\pm\!\!$ &  1.02 & $\!\!\pm\!\!$ &  1.39 &  &  &  \multicolumn{5}{c|}{ } \\
0.24--0.30 & 0.269 &  25.0 &  26.71 & $\!\!\pm\!\!$ &  0.79 & $\!\!\pm\!\!$ &  1.15 & 0.271 &  35.0 &  26.35 & $\!\!\pm\!\!$ &  0.77 & $\!\!\pm\!\!$ &  1.07 \\
0.30--0.36 & 0.329 &  24.9 &  27.13 & $\!\!\pm\!\!$ &  0.79 & $\!\!\pm\!\!$ &  1.06 & 0.330 &  34.9 &  25.06 & $\!\!\pm\!\!$ &  0.75 & $\!\!\pm\!\!$ &  0.93 \\
0.36--0.42 & 0.389 &  25.0 &  24.92 & $\!\!\pm\!\!$ &  0.76 & $\!\!\pm\!\!$ &  0.89 & 0.389 &  35.0 &  23.28 & $\!\!\pm\!\!$ &  0.73 & $\!\!\pm\!\!$ &  0.79 \\
0.42--0.50 & 0.459 &  25.0 &  22.28 & $\!\!\pm\!\!$ &  0.60 & $\!\!\pm\!\!$ &  0.73 & 0.459 &  35.0 &  17.73 & $\!\!\pm\!\!$ &  0.55 & $\!\!\pm\!\!$ &  0.57 \\
0.50--0.60 & 0.547 &  24.9 &  19.55 & $\!\!\pm\!\!$ &  0.50 & $\!\!\pm\!\!$ &  0.65 & 0.548 &  34.8 &  16.55 & $\!\!\pm\!\!$ &  0.48 & $\!\!\pm\!\!$ &  0.54 \\
\hline
 \end{tabular}
 \end{center}
 \end{scriptsize}
 \end{table}

 \begin{table}[h]
 \begin{scriptsize}
 \caption{Double-differential inclusive
  cross-section ${\rm d}^2 \sigma /{\rm d}p{\rm d}\Omega$
  [mb/(GeV/{\it c} sr)] of the production of $\pi^+$'s
  in $\pi^-$ + Be $\rightarrow$ $\pi^+$ + X interactions
  with $-8.0$~GeV/{\it c} beam momentum;
  the first error is statistical, the second systematic; 
 $p_{\rm T}$ in GeV/{\it c}, polar angle $\theta$ in degrees.}
 \label{pip.pimbe8}
 \begin{center}
 \begin{tabular}{|c||c|c|rcrcr||c|c|rcrcr|}
 \hline
   & \multicolumn{7}{c||}{$20<\theta<30$}
  & \multicolumn{7}{c|}{$30<\theta<40$} \\
 \hline
 $p_{\rm T}$ & $\langle p_{\rm T} \rangle$ & $\langle \theta \rangle$
  & \multicolumn{5}{c||}{${\rm d}^2 \sigma /{\rm d}p{\rm d}\Omega$}
  &$\langle p_{\rm T} \rangle$ & $\langle \theta \rangle$
  & \multicolumn{5}{c|}{${\rm d}^2 \sigma /{\rm d}p{\rm d}\Omega$} \\
 \hline
0.10--0.13 & 0.116 &  24.5 &  40.03 & $\!\!\pm\!\!$ &  1.60 & $\!\!\pm\!\!$ &  2.82 & 0.116 &  34.7 &  27.55 & $\!\!\pm\!\!$ &  1.32 & $\!\!\pm\!\!$ &  1.90 \\
0.13--0.16 & 0.146 &  24.6 &  51.88 & $\!\!\pm\!\!$ &  1.74 & $\!\!\pm\!\!$ &  2.87 & 0.145 &  34.8 &  38.40 & $\!\!\pm\!\!$ &  1.48 & $\!\!\pm\!\!$ &  2.07 \\
0.16--0.20 & 0.181 &  24.5 &  58.09 & $\!\!\pm\!\!$ &  1.48 & $\!\!\pm\!\!$ &  2.61 & 0.181 &  34.8 &  40.43 & $\!\!\pm\!\!$ &  1.25 & $\!\!\pm\!\!$ &  1.79 \\
0.20--0.24 & 0.220 &  24.6 &  61.57 & $\!\!\pm\!\!$ &  1.52 & $\!\!\pm\!\!$ &  2.38 & 0.220 &  34.6 &  41.25 & $\!\!\pm\!\!$ &  1.23 & $\!\!\pm\!\!$ &  1.59 \\
0.24--0.30 & 0.270 &  24.7 &  58.68 & $\!\!\pm\!\!$ &  1.19 & $\!\!\pm\!\!$ &  1.94 & 0.269 &  34.7 &  43.31 & $\!\!\pm\!\!$ &  1.02 & $\!\!\pm\!\!$ &  1.42 \\
0.30--0.36 & 0.328 &  24.6 &  53.20 & $\!\!\pm\!\!$ &  1.14 & $\!\!\pm\!\!$ &  1.58 & 0.329 &  34.8 &  36.40 & $\!\!\pm\!\!$ &  0.94 & $\!\!\pm\!\!$ &  1.07 \\
\hline
 \end{tabular}
 \end{center}
 \end{scriptsize}
 \end{table}

 \begin{table}[h]
 \begin{scriptsize}
 \caption{Double-differential inclusive
  cross-section ${\rm d}^2 \sigma /{\rm d}p{\rm d}\Omega$
  [mb/(GeV/{\it c} sr)] of the production of $\pi^-$'s
  in $\pi^-$ + Be $\rightarrow$ $\pi^-$ + X interactions
  with $-8.0$~GeV/{\it c} beam momentum;
  the first error is statistical, the second systematic; 
 $p_{\rm T}$ in GeV/{\it c}, polar angle $\theta$ in degrees.}
 \label{pim.pimbe8}
 \begin{center}
 \begin{tabular}{|c||c|c|rcrcr||c|c|rcrcr|}
 \hline
   & \multicolumn{7}{c||}{$20<\theta<30$}
  & \multicolumn{7}{c|}{$30<\theta<40$} \\
 \hline
 $p_{\rm T}$ & $\langle p_{\rm T} \rangle$ & $\langle \theta \rangle$
  & \multicolumn{5}{c||}{${\rm d}^2 \sigma /{\rm d}p{\rm d}\Omega$}
  &$\langle p_{\rm T} \rangle$ & $\langle \theta \rangle$
  & \multicolumn{5}{c|}{${\rm d}^2 \sigma /{\rm d}p{\rm d}\Omega$} \\
 \hline
0.10--0.13 & 0.116 &  24.8 &  62.88 & $\!\!\pm\!\!$ &  2.02 & $\!\!\pm\!\!$ &  4.24 & 0.116 &  34.8 &  41.80 & $\!\!\pm\!\!$ &  1.61 & $\!\!\pm\!\!$ &  2.90 \\
0.13--0.16 & 0.145 &  24.6 &  85.27 & $\!\!\pm\!\!$ &  2.21 & $\!\!\pm\!\!$ &  4.78 & 0.145 &  34.5 &  53.90 & $\!\!\pm\!\!$ &  1.73 & $\!\!\pm\!\!$ &  2.88 \\
0.16--0.20 & 0.180 &  24.6 &  95.38 & $\!\!\pm\!\!$ &  1.93 & $\!\!\pm\!\!$ &  4.18 & 0.181 &  34.7 &  63.73 & $\!\!\pm\!\!$ &  1.57 & $\!\!\pm\!\!$ &  2.83 \\
0.20--0.24 & 0.220 &  24.7 &  97.33 & $\!\!\pm\!\!$ &  1.93 & $\!\!\pm\!\!$ &  3.65 & 0.220 &  34.7 &  67.33 & $\!\!\pm\!\!$ &  1.59 & $\!\!\pm\!\!$ &  2.57 \\
0.24--0.30 & 0.269 &  24.7 &  96.82 & $\!\!\pm\!\!$ &  1.56 & $\!\!\pm\!\!$ &  3.05 & 0.269 &  34.6 &  62.27 & $\!\!\pm\!\!$ &  1.23 & $\!\!\pm\!\!$ &  1.98 \\
0.30--0.36 & 0.329 &  24.8 &  83.57 & $\!\!\pm\!\!$ &  1.45 & $\!\!\pm\!\!$ &  2.36 & 0.329 &  34.5 &  52.83 & $\!\!\pm\!\!$ &  1.12 & $\!\!\pm\!\!$ &  1.49 \\
\hline
 \end{tabular}
 \end{center}
 \end{scriptsize}
 \end{table}

\clearpage

We refrain from presenting the wealth of cross-section data 
also in the form of plots. We limit ourselves to three 
representative figures that show the inclusive cross-sections
of secondary protons, $\pi^+$'s, and $\pi^-$'s, produced 
by beams of protons, $\pi^+$'s, and $\pi^-$'s. We chose  
a polar-angle range that 
permits a good comparison of our results with 
published results from other experiments. 

\begin{figure}[ht]
\begin{center}    
\includegraphics[width=0.6\textwidth]{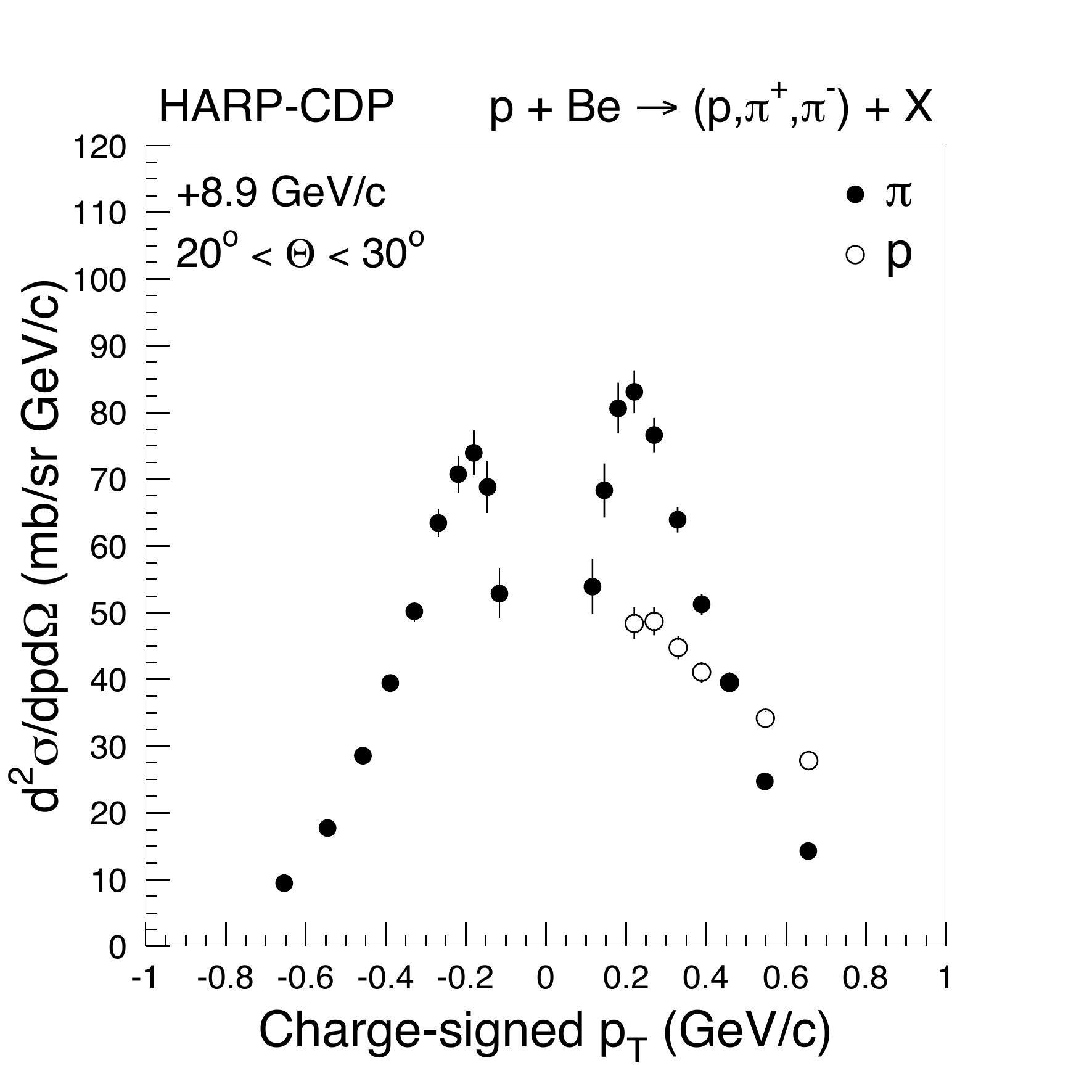}   
\caption{Inclusive cross-sections as a function of 
charge-signed $p_{\rm T}$ 
of proton and $\pi^\pm$ production by  
$+8.9$~GeV/{\it c}
incoming protons, off beryllium nuclei,
in the polar-angle range $20^\circ < \theta < 30^\circ$.}
\label{cross-sectionsfromprotons}
\end{center}
\end{figure}

Figure~\ref{cross-sectionsfromprotons} illustrates our 
measurement of the inclusive  
cross-sections ${\rm d}^2 \sigma / {\rm d} p {\rm d} \Omega$ 
of proton and $\pi^\pm$ production by  
$+8.9$~GeV/{\it c} incoming protons, 
in the polar angle range $20^\circ < \theta < 30^\circ$. 

\begin{figure}[ht]
\begin{center}
\includegraphics[width=0.6\textwidth]{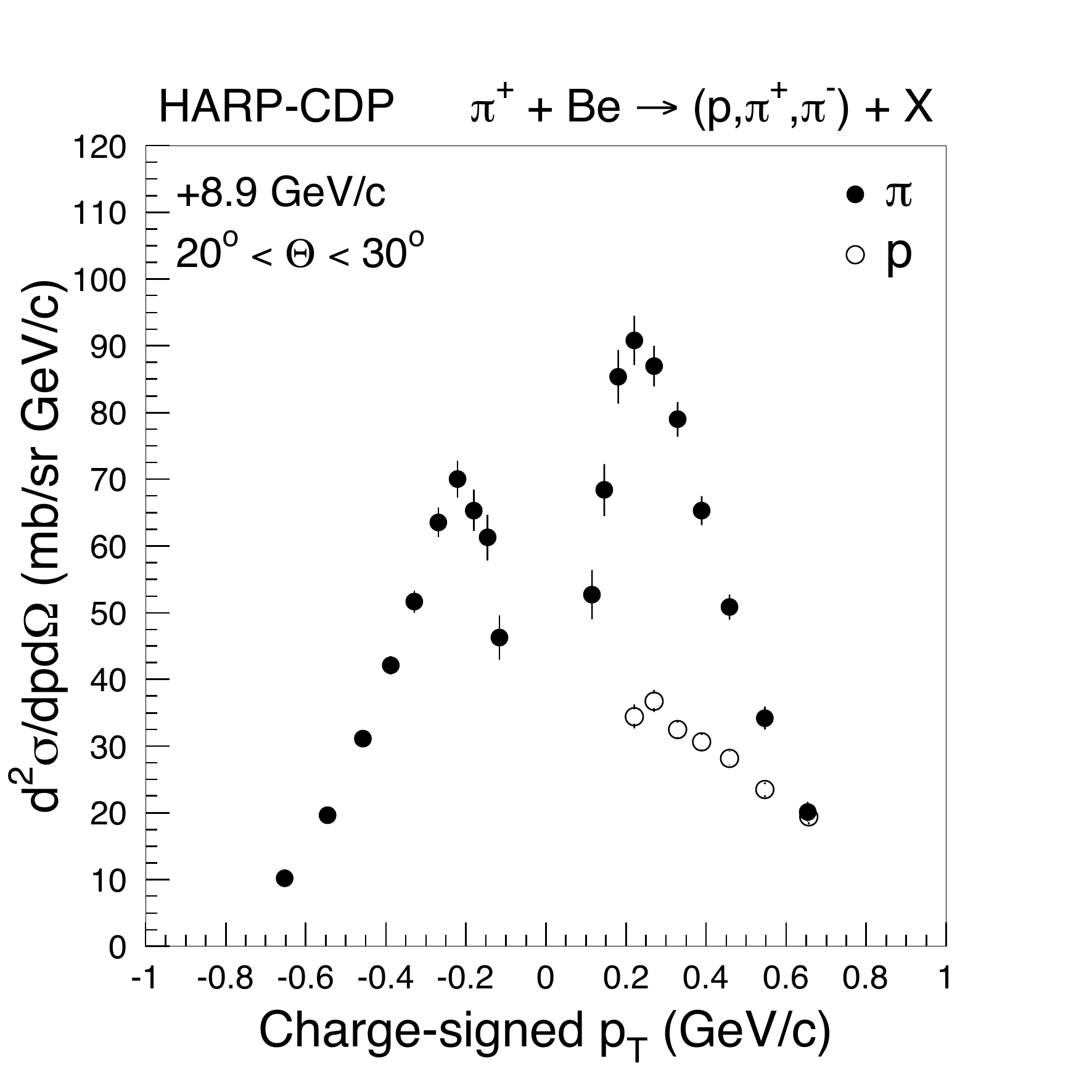} 
\caption{Inclusive cross-sections as a 
function of charge-signed $p_{\rm T}$  
of proton and $\pi^\pm$ production by  
$+8.9$~GeV/{\it c}
incoming $\pi^+$'s, off beryllium nuclei,
in the polar-angle range $20^\circ < \theta < 30^\circ$.}
\label{cross-sectionsfrompiplus}
\end{center}
\end{figure}

Figure~\ref{cross-sectionsfrompiplus} illustrates our 
measurement of the inclusive  
cross-sections ${\rm d}^2 \sigma / {\rm d} p {\rm d} \Omega$ 
of proton and $\pi^\pm$ production by  $+8.9$~GeV/{\it c} incoming $\pi^+$'s, 
in the polar angle range $20^\circ < \theta < 30^\circ$. 

\begin{figure}[ht]
\begin{center}
\includegraphics[width=0.6\textwidth]{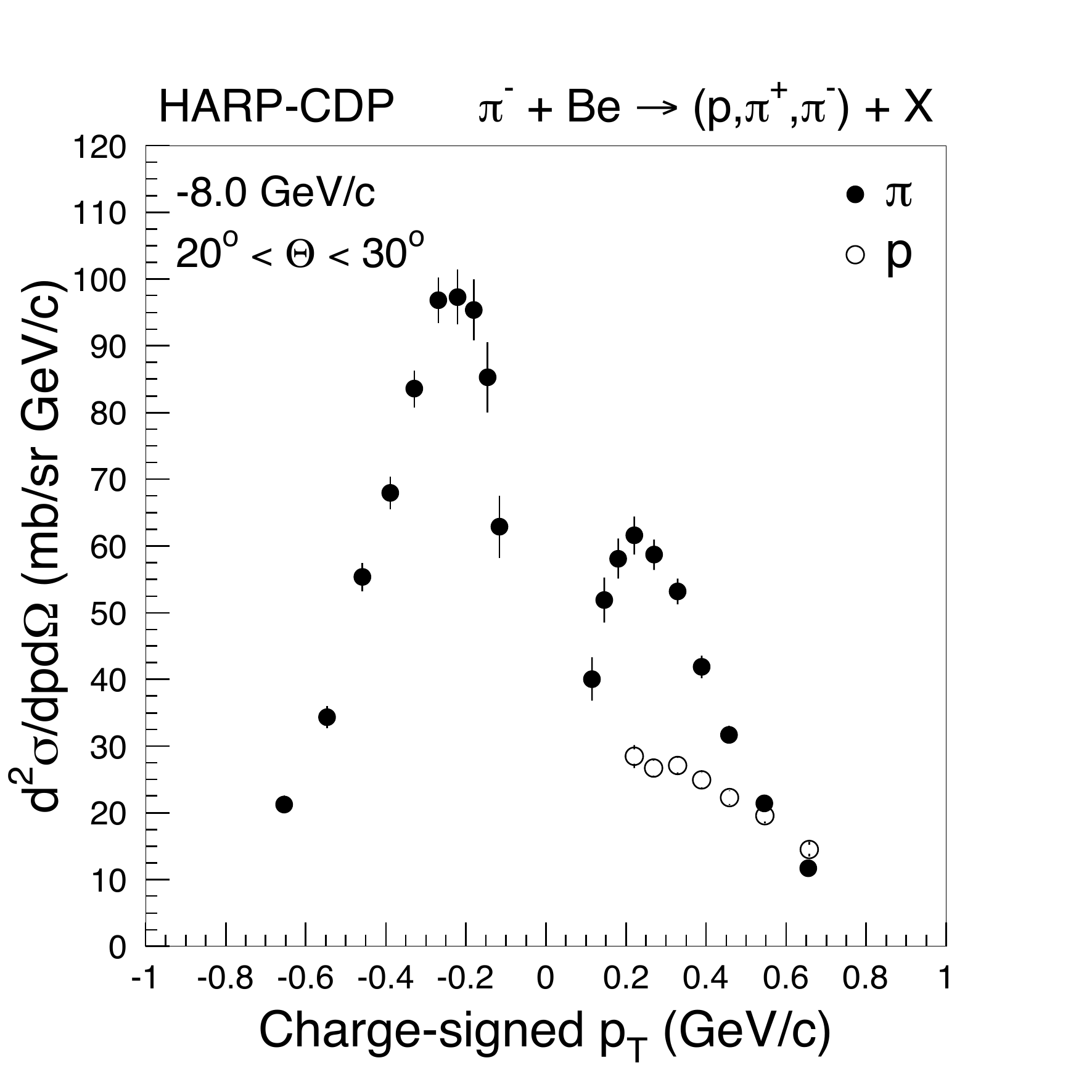} 
\caption{Inclusive cross-sections as a 
function of charge-signed $p_{\rm T}$ 
of proton and $\pi^\pm$ production by $-8.0$~GeV/{\it c}
incoming $\pi^-$'s, off beryllium nuclei,
in the polar-angle range $20^\circ < \theta < 30^\circ$.}
\label{cross-sectionsfrompiminus}
\end{center}
\end{figure}

Figure~\ref{cross-sectionsfrompiminus} illustrates our 
measurement of the inclusive  
cross-sections ${\rm d}^2 \sigma / {\rm d} p {\rm d} \Omega$ 
of proton and $\pi^\pm$ production by $-8.0$~GeV/{\it c} incoming $\pi^-$'s, 
in the polar angle range $20^\circ < \theta < 30^\circ$. 

\clearpage

\section{Comparison with other experimental results}

We compare our $+8.9$~GeV/{\it c} Be cross-sections with those from other experiments and with the results obtained by the HARP Collaboration from the same data that we analysed. The data of the other experiments are from E802 and E910 which were obtained with somewhat higher beam momenta. A more direct comparison with these data will be given in a forthcoming paper where we present cross-sections on Be for all beam momenta from 3 to 15~GeV/{\it c}~\cite{BePaper2}.

\subsection{Comparison with E802 results}
Experiment E802~\cite{E802} at Brookhaven National 
Laboratory measured
secondary $\pi^+$'s in the polar-angle
range $5^\circ < \theta < 58^\circ$ from the interactions of
14.6~GeV/{\it c} protons with beryllium nuclei.

\begin{figure}[ht]
\begin{center}
\includegraphics[width=0.6\textwidth]{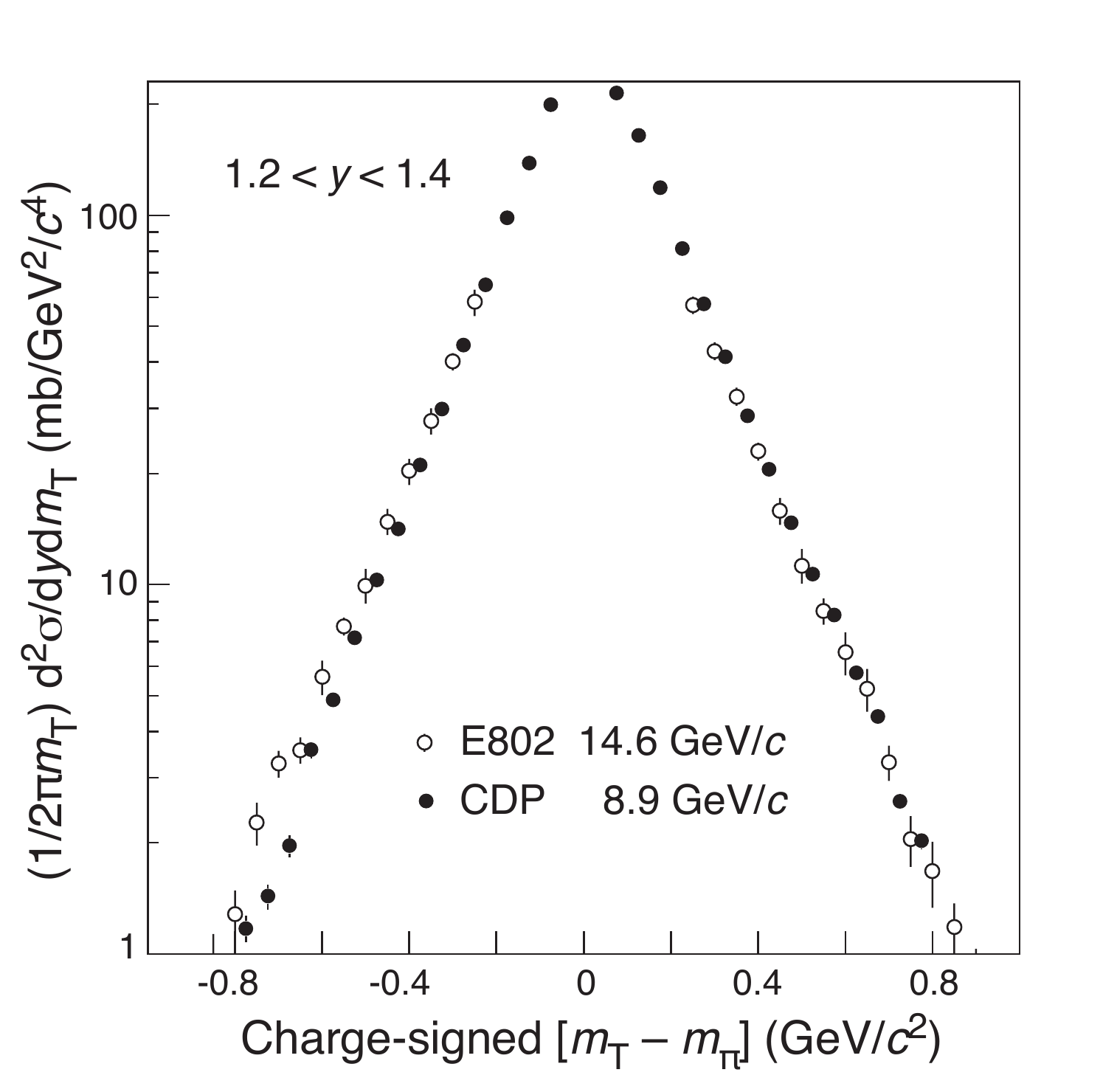}     
\caption{Comparison of our cross-sections (black circles) 
of $\pi^\pm$ production by $+8.9$~GeV/{\it c} 
protons off beryllium nuclei, 
with the cross-sections 
published by the E802 Collaboration for the proton beam 
momentum of 14.6 GeV/{\it c} (open circles); the errors are 
statistical only.}
\label{comparisonwithE802}
\end{center}
\end{figure}

Figure~\ref{comparisonwithE802} shows their published Lorentz-invariant 
cross-section of $\pi^+$ and $\pi^-$ production by
14.6~GeV/{\it c} protons, in the rapidity range $1.2 < y < 1.4$,
as a function of $m_{\rm T} - m_{\pi}$, where $m_{\rm T}$ denotes
the transverse mass. Their data are compared 
with our results expressed in the same units as used by E802. 

We note that both experiments agree in suggesting an exponential
decrease of the invariant cross-section with increasing
$m_{\rm T} - m_{\pi}$, over two orders of magnitude.
Unlike the $\pi^-$ cross-sections, the $\pi^+$ cross-sections 
at $+8.9$ and $+14.6$~GeV/{\it c} exhibit nearly the same slope. In the
comparison of absolute cross-sections, the E802 normalization 
uncertainty of (10--15)\% is to be taken into account on top of 
the beam energy difference.

\subsection{Comparison with E910 results }

Experiment E910~\cite{E910} at Brookhaven National Laboratory
measured secondary charged pions in the momentum
range 0.1--6~GeV/{\it c} from the interactions of 12.3 and 
17.5~GeV/{\it c} protons with beryllium nuclei.
This experiment used a TPC for the measurement of secondaries,
with a comfortably large track length of $\sim$1.5~m. With a magnetic
field strength of 0.5~T, this large
track length renders charge identification and proton--pion 
separation by \dedx\ beyond doubt. 
\begin{figure}[ht]
\begin{center}
\includegraphics[width=0.6\textwidth]{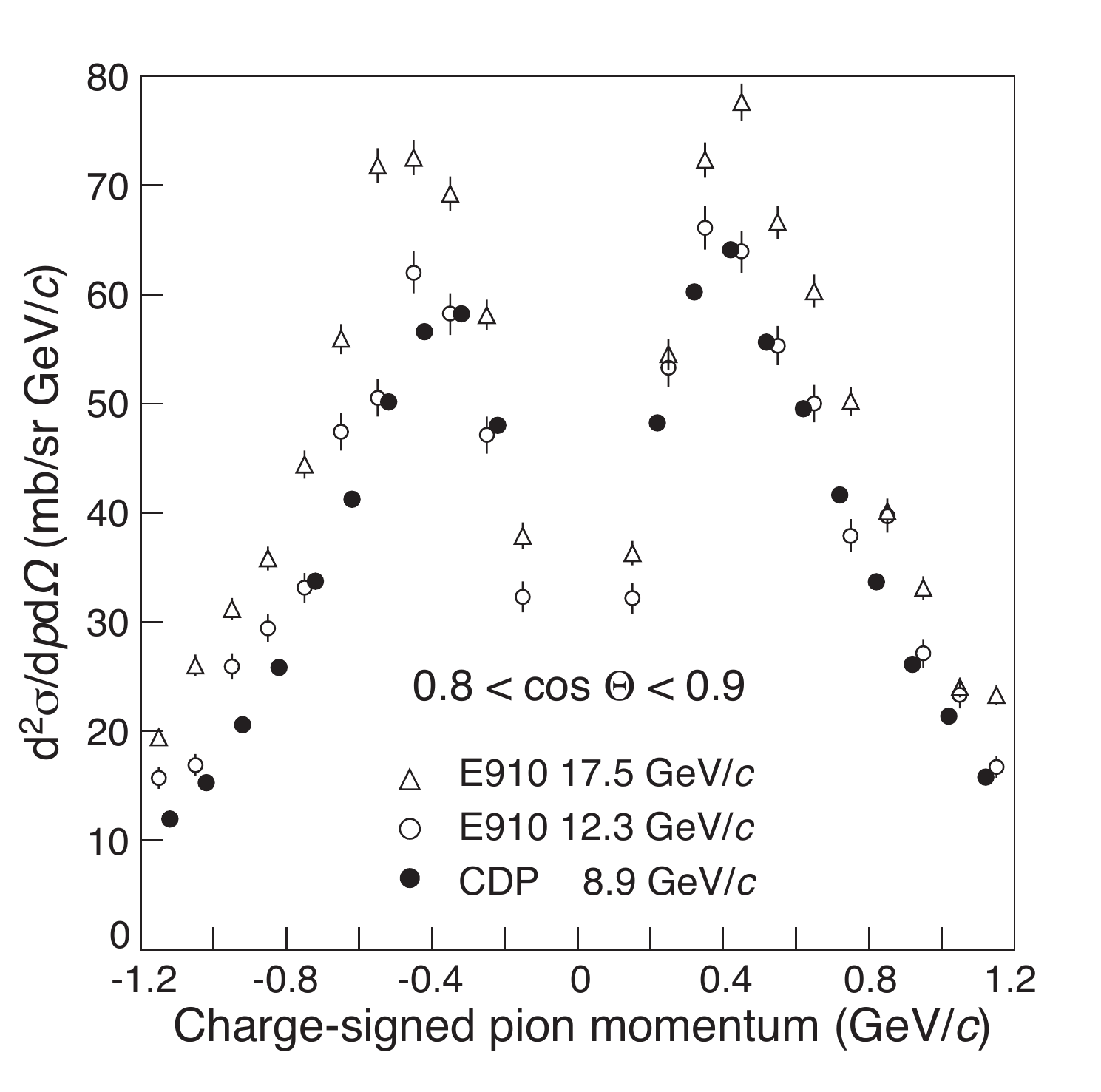}      
\caption{Comparison of our cross-sections (black circles) 
of $\pi^\pm$ production by $+8.9$~GeV/{\it c} 
protons off beryllium nuclei with the cross-sections 
published by the E910 Collaboration for proton beam 
momenta of 12.3 (open circles) and 17.5 (open triangles) 
GeV/{\it c}; the errors are statistical only.}
\label{comparisonwithE910}
\end{center}
\end{figure}

Also here, the E910 data are shown as published, and our data
are expressed in the same units as used by E910. 
Although the E910 measurements were made with proton beam momenta
of 12.3 and 17.5~GeV/{\it c}, respectively, 
we note the similar $\pi^+ / \pi^-$ ratio 
between the cross-sections from E910 and our cross-sections
from a proton beam momentum of 8.9~GeV/{\it c}, shown in 
Fig.~\ref{comparisonwithE910}. In the
comparison of absolute cross-sections, the E910 normalization 
uncertainty of $\leq$5\% is to be taken into account on top of 
the beam energy differences.

\subsection{Comparison with results from the HARP Collaboration}

\begin{figure}[ht]
\begin{center}
\includegraphics[width=0.6\textwidth]{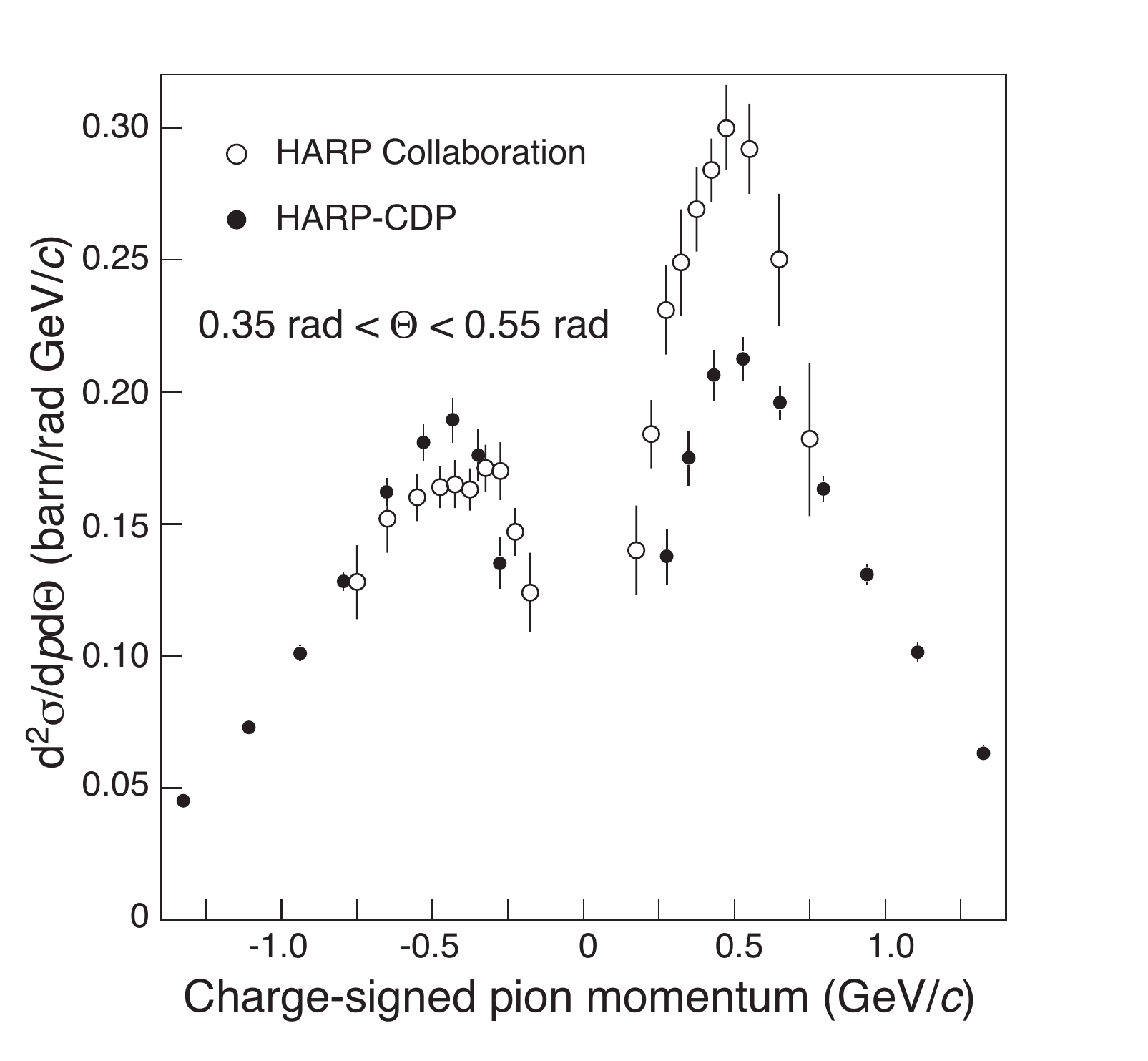}       
\caption{Comparison of our cross-sections (black circles) 
of $\pi^\pm$ production by $+8.9$~GeV/{\it c} 
protons off beryllium nuclei with the cross-sections 
published by the HARP Collaboration (open circles).}
\label{comparisonwithOH}
\end{center}
\end{figure}

Figure~\ref{comparisonwithOH} shows the comparison of our cross-sections of pion production by $+8.9$~GeV/{\it c} protons off beryllium nuclei with the results published by the HARP Collaboration~\cite{OffLApaper}, in the polar-angle range $0.35 < \theta < 0.55$~rad. The latter cross-sections are plotted as published, while we expressed our cross-sections in the units 
used by the HARP Collaboration.

There is a severe discrepancy between our cross-sections and 
those reported by the HARP Collaboration. We note the difference 
especially of the $\pi^+$ cross-section, and the difference in the momentum range. The discrepancy is even more serious as the same data set has been analysed by both groups. For a 
discussion of the reasons for this discrepancy we refer to the Appendix of this paper.

\section{Summary}

From the analysis of data from the HARP large-angle spectrometer
(polar angle $\theta$ in the 
range $20^\circ < \theta < 125^\circ$), double-differential 
cross-sections ${\rm d}^2 \sigma / {\rm d}p {\rm d}\Omega$ 
of the production of secondary protons, $\pi^+$'s, and $\pi^-$'s,
have been obtained. The incoming beam particles 
were $+8.9$~GeV/{\it c} protons 
and pions, and $-8.0$~GeV/{\it c} pions, impinging on
a 5\% $\lambda_{\rm abs}$ thick stationary 
beryllium target. The high statistics of the $+8.9$~GeV/{\it c}
data allowed us to determine cross-sections of K$^+$ and
deuteron production, albeit with lower precision.  
Our cross-sections for $\pi^+$ and $\pi^-$ production 
agree with results from other experiments but disagree with 
the results of the HARP Collaboration that were obtained 
from the same raw data.

\section*{Acknowledgements}

We are greatly indebted to many technical collaborators whose 
diligent and hard work made the HARP detector a well-functioning 
instrument. We thank all HARP colleagues who devoted time and 
effort to the design and construction of the detector, to data taking, 
and to setting up the computing and software infrastructure. 
We express our sincere gratitude to HARP's funding agencies 
for their support.  

\clearpage

\section*{Appendix}

The situation that two groups of authors, in this 
case the `HARP Collaboration' (referred to as `OH' for `Official HARP') and 
us, the HARP--CDP group, publish inconsistent 
results from the same raw data, is unusual and 
unsatisfactory. Naturally, the question arises 
as to whose results can be trusted.

Since OH have not withdrawn their results
despite heavy criticism from independent
review bodies~\cite{CarliFuster,SPSCminutes} and from us,
we cannot avoid addressing this question.

The central problem in OH's data analysis is their 
lack of understanding TPC track distortions which 
leads to:
\begin{itemize}
\item a bias of $\Delta (1/p_{\rm T}) 
\simeq 0.3$~(GeV/{\it c})$^{-1}$ in their reconstruction
of TPC tracks; in other words, their relative $p_{\rm T}$ bias
increases linearly with $p_{\rm T}$ and attains some 30\% at
$p_{\rm T} = 1$~GeV/{\it c}; the bias is such that for 
particles with positive charge $p_{\rm T}$ is decreased, 
while for particles with negative charge $p_{\rm T}$ is increased;
\item a resolution of $\sigma (1/p_{\rm T})  
\simeq 0.6$~(GeV/{\it c})$^{-1}$ which is considerably worse
than $\sigma (1/p_{\rm T}) \simeq$ $0.30$~(GeV/{\it c})$^{-1}$
claimed by OH; and
\item a bad overall RPC time-of-flight resolution of 305~ps and
an apparent advance of the timing signal of protons 
with respect to that of pions by $\sim$500~ps
(`500~ps effect').
\end{itemize}

These three problems, together with a number of 
additional mistakes~\cite{WhiteBook,FurtherWhiteBooks},
have the following fatal consequences 
for cross-sections of secondary 
hadron production:
\begin{itemize}
\item cross-section spectra of secondary hadrons are
distorted especially in regions where cross-sections vary
strongly with momentum; and

\item protons and pions are partly confused.   
\end{itemize}

Discussions of the flaws in the OH analysis have been
published in Refs.~\cite{JINSTpub} and \cite{EPJCpub}.
We summarize here the main arguments. 

OH's argument that `dynamic' TPC track 
distortions can be neglected during the first third of the
400~ms long accelerator spill~\cite{OffTPCcalibration}, 
reads as {\it ``...owing to their
limited mobility the first} [argon] {\it ions created in the amplification 
region need about 25~ms to reach the drift region and
subsequently the steady flow of ions into this region   
only starts approximately 100~ms after the start of the spill,
with a gradual transition between these two regimes...''}
This argument is wrong. With an electric field
strength of $\sim$1.7~kV/cm the argon ions need
less than 1~ms for the relevant 
distance of 11~mm. Therefore, dynamic TPC track distortions
increase right away approximately linearly with time in 
the spill. (Dynamic track distortions in the HARP TPC 
attain in the \rphi\ coordinate $\sim$10~mm at the end
of the spill, one order of magnitude larger than the 
typical \rphi\ resolution; the correct algorithms to cope with
TPC track distortions are described in Ref.~\cite{TPCpub}
and in ample detail in 
Refs.~\cite{distortions,distortions2,distortions3,distortions4}.) 

OH's claims in Ref.~\cite{OffTPCcalibration} that
{\it ``...The constrained fit} [which uses the beam 
point in addition to the TPC clusters] {\it is unbiased with respect 
to the unconstrained fit} [which uses the TPC clusters only]...''
and {\it ``...The weight of the vertex constraint compensates 
very well the distortions...''} contradict logic. It is 
impossible that a circle fit of TPC clusters that shifted away
from their nominal \rphi\ positions, together with the
(undistorted) beam point, yields the same $p_{\rm T}$ as a fit
of the distorted TPC clusters alone. It is equally impossible
that a circle fit of the distorted TPC clusters together 
with the undistorted beam point gives an unbiased $p_{\rm T}$ 
estimate.

OH's justification of their claim of the absence of a
bias in their circle fit stems from Fig.~4 in 
Ref.~\cite{OffTPCcalibration}: {\it ``...In Fig.~4 it is
shown that the vertex constraint does not introduce biases for
those particle trajectories and that the simulation provides an
excellent description of the behaviour of the resolution 
function...''}. Their Fig.~4 claims to prove the 
equality of momentum from OH's `constrained' and 
`unconstrained fits'. That this claim is wrong is evident from 
the unphysical non-Gaussian shape of the shown distribution. 
Its cause is a mistake in their calculation of the 
\rphi\ error of TPC clusters: their
$\sigma^2_{r \! \cdot \! \phi}$ is multiplied by a
factor $\cos^2 {2\phi}$ which assigns clusters an
unphysically large weight depending on how close they are to
the singular values $\phi = 45^\circ$, 135$^\circ$, 
225$^\circ$ and 315$^\circ$ in the azimuthal angle.
(The mathematical intricacies of this mistake are explained 
in Ref.~\cite{WhiteBook}.)  

OH never presented evidence that their 
$p_{\rm T}$ resolution during the accelerator spill is indeed 
$\sigma (1/p_{\rm T}) \simeq$ $0.30$~(GeV/{\it c})$^{-1}$,
and that after TPC track distortion corrections their 
\rphi\ residuals with respect to an unbiased external
coordinated system are compatible with zero across the
whole active TPC volume.

In their most recent physics publication~\cite{OffLApaper}, 
OH claim  
{\it ``...Corrections that allow use of the full statistics 
to be made, correcting for such} [dynamic] {\it distortions, have been
developed...and are fully applied in this analysis. The obtained
results are fully compatible within the statistical errors and
differential systematic uncertainties with those previously
published...''}. This claimed agreement between data from the 
first third of the spill without distortion correction, 
with data from the full spill with distortion correction, 
permits the conclusion that OH's full-spill analysis is beset 
by the same flaws as their earlier analysis of data from the 
first third of the spill.
 
Since OH have a biased track momentum,
they observe that the RPC timing signal of protons is
advanced with respect to the RPC timing signal of 
(relativistic) pions.
This `500~ps effect' observation led them to conclude in 
Ref.~\cite{Version1ofOffTPCcalib} 
{\it ``...While this is in itself an interesting effect
...it prevents the use of the RPCs as a method to verify the
reconstructed momentum scale of heavily ionizing 
particles} [protons].''
As a consequence,  
they made no use of the powerful particle identification 
capability from RPC time of flight. The exclusive use of
\dedx\ from the TPC in conjunction with a biased track 
momentum leads to the partial confusion of protons 
and pions in OH's analysis.  

OH's interpretation of the `500~ps effect' is characterized  
by statements like {\it ``...One possible explanation is 
the fluctuation in arrival time of the first cluster 
of the primary ionization. This fluctuation
is smaller for heavily ionizing particles} [protons]...'' in
Ref.~\cite{500pseffect}, 
or {\it ``...An order of magnitude estimate of the effect
given the propagation velocity of electrons in the gas and
the chamber gap leads to an order of magnitude of a 
few 100~ps...''} in Ref.~\cite{Version1ofOffTPCcalib}.
This understanding of signal generation is wrong.  
The anode signal is generated by induction.
Hence the (fast) propagation of electromagnetic waves 
across the gas gap is relevant and not the arrival at the
anode of---in comparison---slowly moving electrons. (The
correct mechanism of RPC-signal generation is described in 
Ref.~\cite{RPCpub}.)  
 
We hold that results and conclusions published by OH cannot be 
trusted. This refers explicitly to their four physics 
papers~\cite{OffLApaper,OffTapaper,OffCCuSnpaper,OffBeAlPbpaper}, 
four technical papers~\cite{HARPTechnicalPaper,OffRPCPaper,500pseffect,
OffTPCcalibration},
two Rebuttals~\cite{RebuttalTP,RebuttalRPC}, and
one Comment~\cite{CommentTPC}, published to date.

\end{document}